\documentclass[a4paper,fleqn,usenatbib]{mnras}

\usepackage{newtxtext,newtxmath}

\usepackage[T1]{fontenc}
\usepackage{ae,aecompl}

\usepackage{graphicx}	
\usepackage{amsmath}	
\usepackage{amssymb}	
\usepackage{array}

\def\nustar{\textit{NuSTAR}}
\def\xmm{\textit{XMM-Newton}}
\def\xspec{\textsc{xspec}}

\def\nicer{\textit{NICER}}
\def\athena{\textit{ATHENA}}
\def\strobex{\textit{STROBE-X}}

\title[X-ray reverberation model]{A public relativistic transfer function model for
X-ray reverberation mapping of accreting black holes}

\author[A. Ingram et al]{Adam Ingram$^{1}$\thanks{E-mail: adam.ingram@physics.ox.ac.uk},
Guglielmo Mastroserio$^{2}$,
Thomas Dauser $^{3}$,
Pieter Hovenkamp$^{2}$,
\newauthor
Michiel van der Klis$^{2}$
\& Javier A. Garc\'ia$^{4,3}$ \\
$^{1}$Department of Physics, Astrophysics, University of Oxford, Denys
Wilkinson Building, Keble Road, Oxford, OX1 3RH, UK \\
$^{2}$Anton Pannekoek Institute, University of Amsterdam, Science Park
904, 1098 XH Amsterdam, The Netherlands \\
$^3$Remeis Observatory \& ECAP, Universit\"{a}t Erlangen-N\"{u}rnberg,
D-96049 Bamberg, Germany \\
$^4$Cahill Center for Astronomy and Astrophysics, California Institute of Technology, Pasadena, CA 91125, USA}

\date{Accepted 2019 June 18. Received 2019 June 15; in original form 2019 April 26}

\pubyear{2019}

\begin{document}
\label{firstpage}
\pagerange{\pageref{firstpage}--\pageref{lastpage}}
\maketitle

\begin{abstract}
We present the publicly available model
\textsc{reltrans} that calculates the
light-crossing delays and energy shifts experienced by X-ray photons
originally emitted close to the black hole when they reflect from the
accretion disk and are scattered into our line-of-sight, accounting
for all general relativistic effects. Our model is fast and flexible
enough to be simultaneously fit to the observed energy-dependent
cross-spectrum for a large range of Fourier frequencies, as well as to
the time-averaged spectrum. This not only enables better geometric
constraints than only modelling the relativistically broadened
reflection features in the time-averaged spectrum, but additionally
enables constraints on the mass of supermassive black holes in active
galactic nuclei and stellar-mass black holes in X-ray binaries. We
include a self-consistently calculated radial profile of the
disk ionization parameter and properly account for the effect that the
telescope response has on the predicted time lags. We find that a
number of previous spectral analyses have measured artificially low
source heights due to not accounting for the former effect and that
timing analyses have been affected by the latter. In particular, the
magnitude of the soft lags in active galactic nuclei may have been
under-estimated, and the magnitude of lags attributed to thermal
reverberation in X-ray binaries may have been over-estimated. We fit
\textsc{reltrans} to the lag-energy spectrum of the Seyfert galaxy Mrk
335, resulting in a best fitting black hole mass that is smaller than
previous optical reverberation measurements ($\sim 7$ million compared
with $\sim14-26$ million $M_\odot$).
\end{abstract}




\section{Introduction}
\label{sec:intro}

Stellar-mass black holes in X-ray binary systems and supermassive
black holes in active galactic nuclei (AGN) are thought to accrete via a
geometrically thin, optically thick accretion disk, which radiates
thermally (\citealt{Shakura1973,Novikov1973}). The hard X-ray spectrum
is often dominated by a power-law component with a high energy cut
off, thought to be due to Compton up-scattering of comparatively cool
photons in a cloud (with optical depth $\tau\sim1-2$) of hot electrons
located close to  the black hole
(\citealt{Thorne1975,Sunyaev1979}). The exact geometry of this cloud
is still debated, with suggested models including a standing shock at
the base of the jet (\citealt{Miyamoto1991,Fender1999}), a coronal
layer sandwiching the disk (\citealt{Galeev1979,Haardt1991}), and
evaporation of the inner disk regions to form a hot, large scale
height accretion flow (\textit{the truncated disk model};
\citealt{Eardley1975,Ichimaru1977,Done2007}). In the absence of a
consensus on its geometry, the Comptonising region is often simply
referred to as the corona, a convention that we will employ here.

We observe Comptonized radiation that reaches us directly from the
corona (the \textit{direct} component) in addition to coronal emission
that has been reprocessed and re-emitted in the upper atmosphere of
the disk, conventionally called the \textit{reflection}
component. These `reflected' photons imprint characteristic features
onto the observed spectrum including a prominent iron K$\alpha$
emission line at $\sim 6.4$ keV and a so-called reflection hump
peaking at $\sim20-30$ keV
(e.g. \citealt{Lightman1980,George1991,Ross1993,Garcia2010}). The iron
line provides a powerful probe of the disk dynamics and geometry,
since its shape is observed to be distorted by photon energy shifts
caused by relativistic orbital motion of disk material and
gravitational redshift (\citealt{Fabian1989,Laor1991}). If the disk
inner radius inferred from the line profile is sufficiently small,
setting it equal to the innermost stable circular orbit (ISCO) of
general relativity (GR) provides an estimate for the spin of the black
hole.

Many studies have used reflection spectroscopy to probe both AGN
(e.g. \citealt{Tanaka1995,Reynolds2003,Patrick2012,Walton2013,Risaliti2013}) 
and black hole X-ray binary
(e.g. \citealt{Miller2007,Reis2009a,Miller2013,Kolehmainen2014,Plant2014,Garcia2015})
accretion flows. This has yielded many measurements of high black hole
spin in AGN (e.g. \citealt{Reynolds2019,Middleton2016a}), although complex line-of-sight
absorption can potentially introduce modelling systematics
(e.g. \citealt{LMiller2008}). For the binaries, spectral modelling
studies often conclude that the inner radius moves towards the black
hole as the spectrum evolves from the hard power-law dominated
\textit{hard state} to the thermal disk dominated \textit{soft state}
on timescales of $\sim$months
(\citealt{Done2007,Plant2014,Garcia2015}). However, even though there
is broad agreement in the \textit{trend} in disk inner radius, the
measured \textit{values} themselves vary enormously between different
studies (\citealt{Garcia2015}), with potential systematics including
calibration uncertainty (e.g. \citealt{Done2010}) and the difficulty
of disentangling the direct and reflected components
(e.g. \citealt{Basak2017}).

The degeneracies associated with spectral modelling can be addressed
by additionally modelling the light-crossing delay between variations
in the direct and reflected spectral components 
(\citealt{Campana1995,Reynolds1999,Uttley2014}). Such
reverberation mapping techniques therefore promise better
constraints on the disk geometry (and therefore the black hole spin),
but also entirely new constraints on black hole mass (\citealt{Stella1990}).
This is essentially because the delays depend on physical distances,
whereas the energy shifts are only sensitive to distances in units of
gravitational radii ($R_g=GM/c^2$). Fourier frequency dependent time
lags between energy channels can be calculated from the argument of
the cross-spectrum (\citealt{vanderKlis1987}). It is routinely found
that, at low Fourier frequencies ($\nu \lesssim 1.5\times 10^{-3}
c/R_g$), hard photons lag soft, both for the binaries
(\citealt{Miyamoto1988,Nowak1999,Kotov2001}) and AGN 
(e.g. \citealt{Papadakis2001,McHardy2004,Epitropakis2017}). These
\textit{intrinsic hard lags} are thought to be caused by spectral
variability of the direct component rather than reverberation, due to
their large magnitude and the lack of reflection features in their
energy dependence, and may originate from inward propagation of
fluctuations in the mass accretion rate
(\citealt{Arevalo2006,Ingram2013,Rapisarda2017a,Mahmoud2018}). Since 
the hard lags reduce with frequency, it has been possible to detect
reverberation signatures at high frequencies in AGN, first through
soft lags interpreted as the soft-excess of the reflection spectrum
lagging the direct radiation (\citealt{Fabian2009,DeMarco2013}), and later
through an iron line feature in the lag-energy spectrum
(\citealt{Zoghbi2012,Kara2016}). A number of studies have focused on
modelling these high frequency lags in AGN
(\citealt{Cackett2014,Emmanoulopoulos2014,Epitropakis2016,Chainakun2016,Wilkins2016,Caballero-Garcia2018}).
Discoveries of reverberation signals came a little later for the
binaries, since the cross over from intrinsic to reverberation lags is
at a much higher frequency (measured in Hz rather than $c/R_g$) for a
stellar-mass black hole (due to mass scaling), and therefore the
signal is harder to pick out of the Poisson noise. Still, soft lags,
interpreted as reverberation of thermally reprocessed photons, have
been detected (\citealt{Uttley2011,DeMarco2015} - although this could
feasibly result from backwards propagation of accretion rate
fluctuations: \citealt{Mushtukov2018}), and iron K lags were finally
detected for MAXI J1820+070 by \cite{Kara2019} using the
\textit{Neutron star Interior Composition ExploreR}
\citep[\textit{NICER};][]{Gendreau2016}.

Still further information is contained in the energy and frequency
dependent variability amplitude, which can also be measured from the
cross-spectrum. This can provide powerful constraints, since the
variability amplitude of reflected emission should be washed out at
the highest frequencies due to destructive interference between rays
reflected from different parts of the disk
(\citealt{Gilfanov2000}). It is optimal to consider all of these
properties simultaneously\footnote{Indeed, also considering the power
  spectrum additionally provides information on the coherence between
  energy bands (e.g. \citealt{Rapisarda2016}), which we ignore
  here.}. The neatest way to do this statistically is to jointly model
the time-averaged spectrum and the real and imaginary parts of the
energy dependent cross-spectrum (\citealt{Mastroserio2018}). Here we
present a public \textsc{xspec} model that enables such an
analysis. We define two versions of the model, \textsc{reltrans} and
\textsc{reltransCp}, which represent the direct spectral component
respectively as an exponentially cut-off power-law and using the model
\textsc{nthcomp}. We assume a simple lamppost geometry
(\citealt{Matt1991,Martocchia1996}), which allows all GR effects to be
properly accounted for without prohibitive computational expense -
although we note that it is simple in our formalism to consider a
number of lamppost sources. Source code and usage instructions can be
downloaded from \url{https://adingram.bitbucket.io/}.

We present a detailed derivation of our model in Section
\ref{sec:deriv} and explore its properties in Section
\ref{sec:results}. Our treatment properly accounts for line-of-sight
absorption and the telescope response, and the model accounts for the
radial dependence of the  disk ionization parameter. In Section
\ref{sec:bias} we investigate the importance of these effects.
In Section \ref{sec:Mrk335}, we perform a proof-of-principle fit to
the lag-energy spectrum of the narrow-line Seyfert 1 galaxy Mrk 335
for a single frequency range. We discuss and conclude our findings in
Sections \ref{sec:discussion} and \ref{sec:conclusions}.

\section{Derivation of the cross-spectrum in the lamppost geometry}
\label{sec:deriv}

\begin{figure}
	\includegraphics[width=8cm,trim=2.0cm 0.5cm
        2.0cm 0.0cm,clip=true]{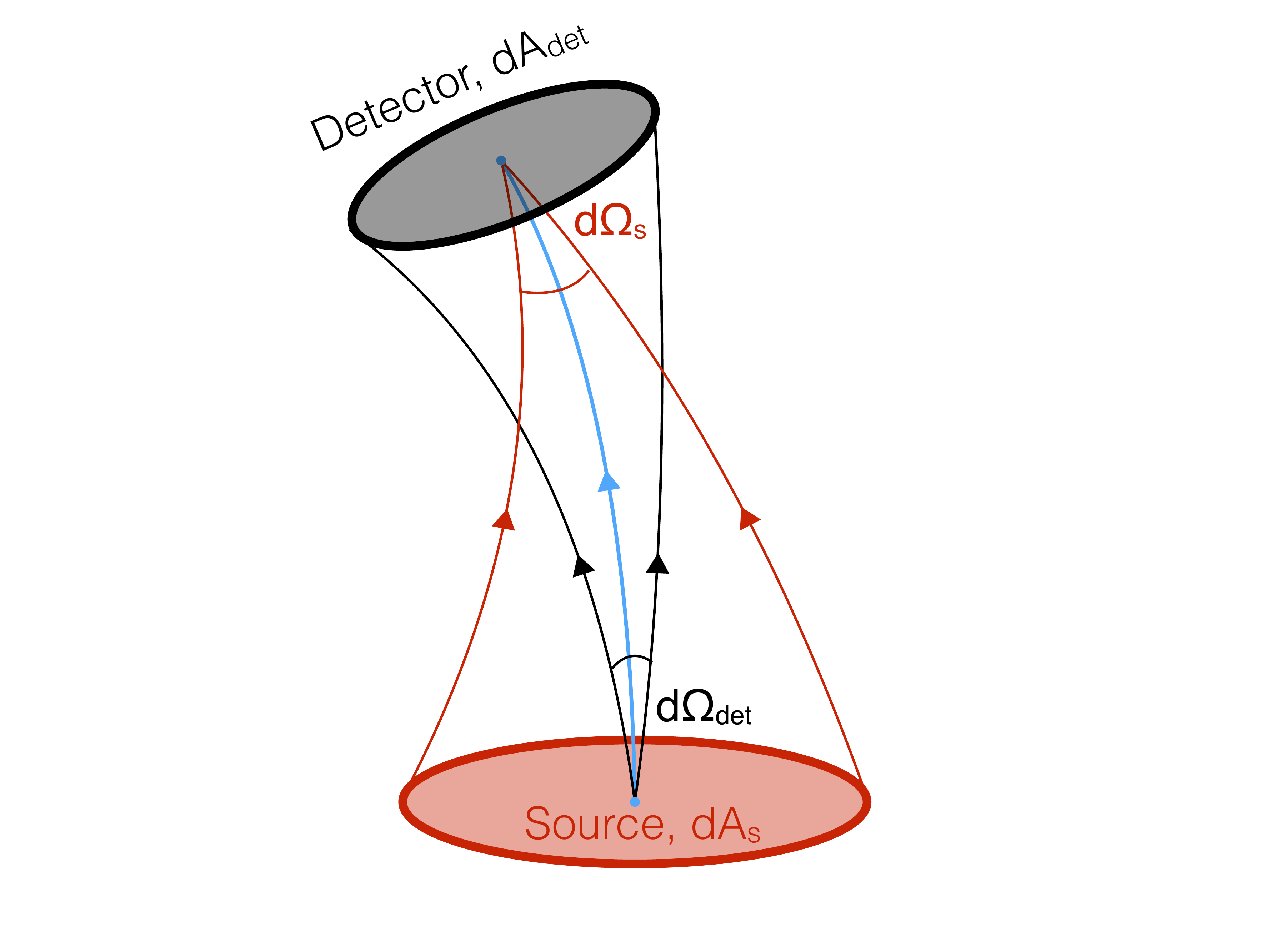} 
\vspace{0mm}
\caption{Schematic of a source and detector with surface areas
  (measured in their own restframes) $dA_s$ and $dA_{\rm det}$
  respectively. The blue line represents a photon path that emerges
  parallel to the source surface area vector (in the source restframe)
  and arrives parallel to the detector surface area vector (in the
  detector restframe). Only photons emerging from the source within
  the solid angle $d\Omega_{\rm det}$ will eventually hit the detector. The solid angles
  and surface areas are related through the reciprocity theorem
  (equation \ref{eqn:reciprocity}).}
 \label{fig:solidangle}
\end{figure}

\begin{figure*}
	\includegraphics[width=17cm,trim=2.3cm 9.5cm
        2.0cm 7.0cm,clip=true]{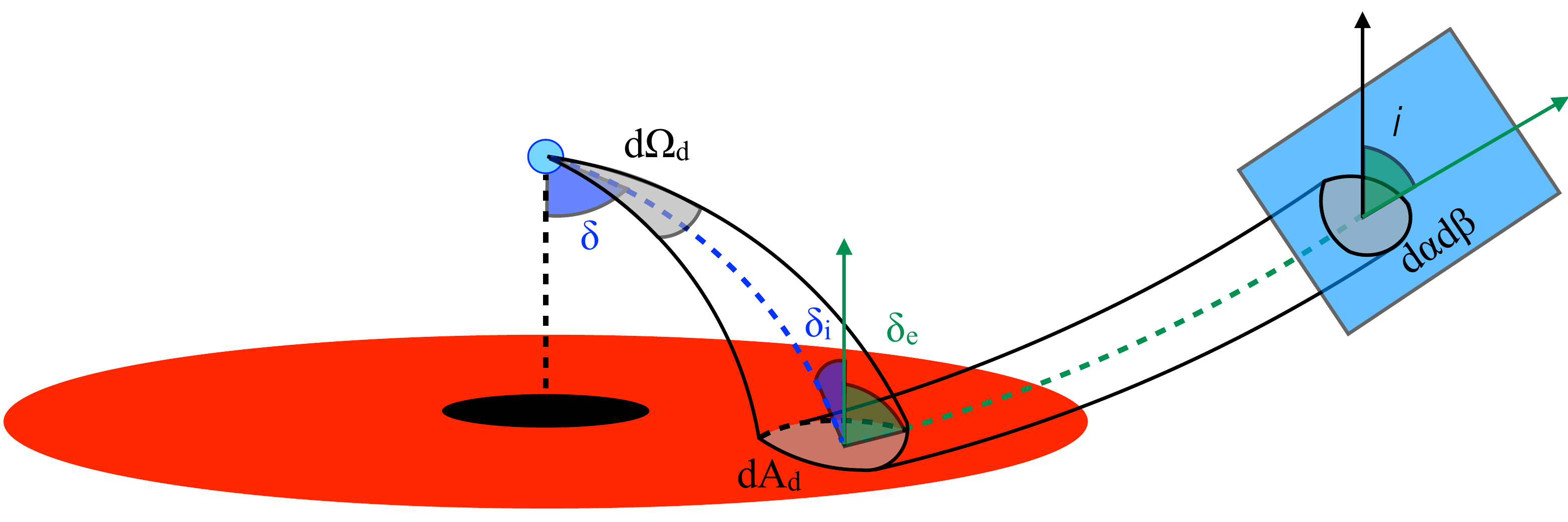} 
\vspace{0mm}
 \caption{Schematic of the on-axis lamppost geometry. A disk patch
   with area $dA_d$ subtends a solid angle $d\Omega_d$ according to
   the irradiating source. The disk patch corresponds to an area
   $d\alpha d\beta$ on the image plane, where $\alpha$ and $\beta$ are
   respectively horizontal and vertical impact parameters at
   infinity. The bundle of rays within the represented solid angle are
   assumed to follow the trajectory (green dashed lines) defined by
   the initial ($\delta$), incidence ($\delta_i$), emission
   ($\delta_e$) and inclination ($i$) angles.}
 \label{fig:schem}
\end{figure*}

Here we derive the time-dependent observed energy spectrum
assuming an isotropically radiating lamppost source located on the
black hole spin axis a height $h$ above the hole, and use the Fourier
transform to calculate the energy dependent cross-spectrum as
a function of Fourier frequency $\nu$. We assume that the specific
(energy) flux (i.e. energy per unit time per unit area per
unit photon energy) seen by a distant observer as a function of photon
energy, $E$, and time $t$, both defined in the observer's restframe,
is given by
\begin{equation}
S(E,t) = F(E,t) + R(E,t).
\end{equation}
The first and second terms on the right hand side represent
respectively the direct and reflected spectral components. In this
paper, we ignore directly observed disk radiation, assuming
it to be below our X-ray bandpass. This is appropriate for AGN, and
hard state X-ray binaries in the $E\gtrsim 3$ keV bandpass.

In this section, we first go over some general considerations of
radiation theory in GR (Section \ref{sec:rec}). We then derive the observed
time-dependent direct spectrum (Section \ref{sec:cont}) and the
reflection spectrum (Section \ref{sec:ref}), before deriving the
transfer function that we will use for our reverberation model
(Section \ref{sec:trans}), followed by the kernel used to calculate
our transfer function (Section \ref{sec:kernel}). Finally we will
discuss the so-called reflection fraction (Section
\ref{sec:reflfrac}).

\subsection{Reciprocity and Liouville's theorem}
\label{sec:rec}

Fig \ref{fig:solidangle} shows a schematic of a source with surface
area $dA_s$ in its own rest frame and a detector with surface area
$dA_{det}$ in its own rest frame. Photons travel along null
geodesics, which are solutions to the geodesic equation with line
element $ds^2 = g_{\mu\nu}dx^\mu dx^\nu=0$. Here, $g_{\mu\nu}$ is the
metric and $dx^\mu$ is the coordinate interval corresponding to an
interval $ds/c$ in proper time. Throughout this paper, we use the Kerr
metric in Boyer-Lindquist coordinates. The position of a photon along
its geodesic is described by the affine parameter $\lambda$ and its
trajectory described by the tangent vector
$k^\mu(\lambda)=dx^\mu/d\lambda$. The blue line in Fig
\ref{fig:solidangle} represents the unique null-geodesic,
$k^\mu(\lambda)$, that connects the centre of the source to the centre
of the detector. For this example, the geodesic begins parallel to the
source's surface area vector and ends parallel to the detector's
surface area vector, but we can generalize by specifying $dA_{det}$
and $dA_s$ to be respectively the \textit{projected} area of the
detector and source perpendicular to $k^\mu$ in the local
restframe. The black lines depict the trajectory of photons that
emerge from the centre of the source and hit the edge of the detector,
representing a bundle of photons that diverge from the centre of the
source around $k^\mu(\lambda)$, subtending a solid angle
$d\Omega_{det}$ in the source rest frame (i.e. all the photons in the
bundle hit the detector and all others miss). The red lines depict the
trajectory of photons that emerge from the edge of the source and hit
the centre of the detector, representing a bundle of geodesics that
converge onto the centre of the detector around $k^\mu(\lambda)$, 
subtending a solid angle $d\Omega_s$ in the detector rest frame. This
is the solid angle that the source subtends on the detector's sky
(i.e. all photons from the bundle hit the centre of the detector and
all others miss the centre).

These four quantities are related by the general relativistic
\textit{reciprocity theorem}
\begin{equation}
  g_{so}^2 dA_{det} d\Omega_s = dA_s d\Omega_{det}. \label{eqn:reciprocity}
\end{equation}
Here $g_{so}=E_{o}/E_s$ is the blueshift\footnote{i.e. $g_{so}>1$
  corresponds to a blueshift and $g_{so}<1$ to a redshift. Note that
  $g_{so}=1/(1+z_{so})$, where $z_{so}$ is redshift defined in the
  traditional sense of fractional change in wavelength.} experienced
by photons traveling 
from source to detector and $E_{o}$ and $E_s$ are respectively the
energy of the photon as measured in the rest frame of the detector and
source (see Appendix \ref{sec:gandmu} for expressions of blueshift
factors in the Kerr metric). The reciprocity theorem in GR was first derived by
\cite{Etherington1933}, and a more concise presentation of the
derivation can be found on pages 631-633 of \citet[this is a
republication of the original 1971
proceeding]{Ellis1971}. \cite{Ellis2007} provides a useful commentary
on the original Etherington paper. This is an intriguing geometrical
result, showing that the curvature of spacetime does not influence the
relationship between these solid angles and surface areas: the
reciprocity theorem in GR is the same as that in special relativity
for a given blueshift factor. We can even recover the classical
reciprocity theorem by transforming $d\Omega_{det}$ into the detector
frame to cancel out the  $g_{so}^2$. The blueshift is calculated as
\begin{equation}
g_{so} = \frac{ (k_{det})^\mu (u_{det})_\mu  }{
  (k_s)^\nu (u_s)_\nu},
\label{eqn:gso}
\end{equation}
where $u_s$ and $u_{det}$ are respectively the 4-velocity of the
source and detector.

If $dN(E_s)$ photons with energies ranging from $E_s$ to $E_s+dE_s$
are radiated isotropically from the flat source surface in its
restframe, a fraction $d\Omega_{det}/\pi$ of them will cross the
detector some time later\footnote{This is because the projected area
  of the source is $\propto\cos\theta$ if it is viewed from inclination
  $\theta$ (and $d\Omega=d\cos\theta d\phi$). Therefore the
  fraction is $d\Omega/\int_0^{2\pi} \int_0^1 \cos\theta
  d\cos\theta d\phi = d\Omega/\pi$}. Their energies will be
measured to range from $E_o$ to $E_o+dE_o$ in the detector rest
frame, meaning that the number of photons crossing the detector is
$dN_o(E_o)=dN(E_s) d\Omega_{det}/\pi$. The specific (energy) flux
crossing the detector is therefore
\begin{eqnarray}
F_{o}(E_o,t_o) &\equiv& \frac{E_o dN_o(E_o)}{dt_o dE_o dA_{det}}
                     \nonumber \\
&=& \frac{g_{so}}{\pi} \frac{E_s dN(E_s)}{dt_s dE_s dA_{s}} dA_s
    \frac{d\Omega_{det}}{dA_{det}} \nonumber \\
 &=& \frac{g_{so}}{\pi} F_s(E_s) dA_s \frac{d\Omega_{det}}{dA_{det}},
     \label{eqn:fgen}
\end{eqnarray}
where $F_s(E_s) $ is the specific flux radiated by the source,
$dt_s = g_{so} dt_o$ and we have used $dE_o/dE_s =
g_{so}$. Rearranging the above equation and applying the reciprocity
theorem (equation \ref{eqn:reciprocity}) gives the rather familiar
formula
\begin{equation}
I_o(E_o) = g_{so}^3 I_s(E_s),
\label{eqn:liouville}
\end{equation}
where $I_o(E_o)$ and $I_s(E_s)$ are specific intensities: specific
flux per unit solid angle (in this case $I_o(E_o)=F_o(E_o)/d\Omega_s$ and
$I_s(E_s)=F_s(E_s)/\pi$). This famous result can also be derived from
Liouville's theorem, which states that the number of photons per unit
volume in phase space is Lorentz invariant (see
e.g. \citealt{Lindquist1966,Misner1973}). The derivation presented
here is perhaps more intuitive. Integrating both sides over all
observed energies gives the familiar expression for bolometric flux in
terms of bolometric intensity
\begin{equation}
F_{o} = g_{so}^4 I_s d\Omega_s.
\end{equation}
We can understand intuitively where these four factors of blue shift
originate. Two come from the adjustment to solid angle in the
reciprocity theorem (equation \ref{eqn:reciprocity} - and these two
factors can further be understood as special relativistic aberration
in the small angle limit), one comes from the adjustment to the energy
of each photon and one comes from the adjustment to time intervals
(i.e. bolometric flux is $\propto E dN/dt$). Finally, all blue shifts
in this paper are calculated in the Kerr metric, which is
asymptotically flat and stationary and therefore does not account for
cosmological redshift. An observer at cosmological redshift $z$ will
therefore measure a specific intensity
\begin{equation}
I(E) = \left(\frac{g_{so}}{1+z}\right)^3 I_s(E_s),
\label{eqn:zliouville}
\end{equation}
and will measure time intervals $\tau = (1+z) \tau_o$ (cosmological
time dilation).

\subsection{Direct spectrum}
\label{sec:cont}

We assume a spherical X-ray source, with surface area in its own rest
frame $a_s$, that isotropically radiates a specific flux
\begin{equation}
F_s(E_s,t_o') = \frac{C(t_o')}{a_s} f(E_s|\Gamma,E_{cut}),
\label{eqn:Fs}
\end{equation}
where $t'_o$ is time in the restframe of an observer at cosmological
redshift $z=0$. In the \textsc{reltrans} version of the model, the
direct spectrum is
\begin{equation}
f(E|\Gamma,E_{cut}) \propto E^{1-\Gamma} {\rm e}^{-E/E_{cut}},
\end{equation}
where the constant of proportionality will be calculated below. The
\textsc{reltransCp} version instead uses the thermal Comptonisation
model \textsc{nthcomp} (\citealt{Zdziarski1996,Zycki1999}), with the
$E_{cut}$ parameter replaced by the electron temperature $kT_e$. For
the purposes of this derivation, we will always use $E_{cut}$, on the
understanding that this can be replaced with $kT_e$ for the case of
the \textsc{nthcomp} version. In order to evaluate the function $f(E)$
in these two cases, we use the model \textsc{xillver} and
\textsc{xillverCp} respectively (\citealt{Garcia2010,Garcia2013}),
which we will also use in order to calculate the restframe reflection
spectrum. Our code calls the relevant \textsc{xillver} model with the
reflection fraction parameter set to zero, which returns the
illuminating spectrum used for the calculation of the reflection
spectrum.

In this paper, as can be seen in Equation (\ref{eqn:Fs}), we will only
consider \textit{linear} variability of the source flux. That is, the
shape of the direct component of the spectrum remains constant in time
and only the normalisation varies. In future versions, we will extend our
modelling to account for non-linear variations of the spectrum
radiated by the corona using the Taylor expansion technique described
in \cite{Mastroserio2018}.

We assume the source is small enough to ensure that any light rays
that pass by either side of it on route to the distant observer are
parallel to one another (i.e. spacetime is approximately flat on the
scale of the source area). The projected area of the source is
therefore $dA_s=a_s/4$. Substituting this into equation \ref{eqn:fgen}
gives the observed specific flux at time $t_o$
\begin{eqnarray}
 F_{o}(E_o,t_o) &=& \frac{g_{so}}{4\pi} F_s(E_s,t_o') a_s
                   \frac{d\Omega_{det}}{dA_{det}} \nonumber \\
&=& \frac{C(t_o')}{4\pi} g_{so}^\Gamma f(E_o|\Gamma,g_{so}E_{cut})
            \frac{d\Omega_{det}}{dA_{det}},
	\label{eqn:Fso1}
\end{eqnarray}
where $t_o=t_o'+\tau_{so}$ and $\tau_{so}$ is the time it takes
photons to travel from source to detector, as measured in the detector
frame (and assuming $z=0$). The second line of the above equation is
exact for an exponentially cut-off power-law illuminating spectrum but
only approximate for an \textsc{nthcomp} spectrum. The final term on the
right hand side accounts for lensing / de-lensing due to light
bending. Defining the inclination angle $i$ as the angle between the
black hole spin axis and the trajectory of photons when they cross the
detector (see Fig \ref{fig:schem}, but note that photons reach the
detector both directly and via reflection) and
$D=\sqrt{dA_s/d\Omega_s}$ as the distance between the source and the
detector (and also, to a very good approximation, the distance between
the hole and the detector, since $D>>h$), the detector area is
$dA_{det}=D^2 \sin i~di~d\phi$. Defining $\delta$ as the angle,
measured in the source rest frame, between the spin axis and the
emergent trajectory of a photon as it is radiated by the source (see
Fig \ref{fig:schem} for an example of a photon that reflects from the
disk, but note that photons with larger $\delta$ may reach the
observer directly), we can write
\begin{equation}
\frac{d\Omega_{det}}{dA_{det}} = \frac{1}{D^2} 
  \left| \frac{d\cos\delta}{d\cos i} \right|  =
\frac{\ell}{D^2},
\end{equation}
since intervals in azimuth $d\phi$ are constant along a geodesic for
an on-axis source in the Kerr metric. We calculate the lensing factor,
$\ell=|d\cos\delta/d\cos i|$, numerically by tracing rays along null
geodesics in the Kerr metric, calculated using the publicly available
code \textsc{ynogk} (\citealt{ynogk}), which is based on another
publicly available code \textsc{geokerr} (\citealt{Dexter2009}). The
observed specific flux is therefore
\begin{equation}
F_{o}(E_o,t_o) = A(t_o') \ell g_{so}^\Gamma f(E_o|\Gamma,g_{so}E_{cut}),
\label{eqn:Fso}
\end{equation}
where we have defined $A(t) \equiv C(t) / ( 4\pi D^2 )$. An observer
at a cosmological distance sees a specific flux
\begin{equation}
F(E,t) = A(t') \ell \left( \frac{g_{so}}{1+z} \right)^\Gamma
f[E|\Gamma,g_{so}E_{cut}/(1+z)],
\label{eqn:F}
\end{equation}
and measures a time interval $t-t'=(1+z) \tau_{so}$. In our model, for
consistency with the \textsc{relxill} family of models
(\citealt{Dauser2013,Garcia2014}), we specify as a model parameter the
cut-off energy in the \textit{observer's} frame, $(E_{cut})_{obs} =
g_{so} E_{cut} / (1+z)$. When the verbose level is set suitably high,
the code prints to screen the value of the cut-off energy in the
source restframe. For \textsc{reltransCp}, we instead specify the
parameter $(kT_e)_{obs} = g_{so} kT_e/ (1+z)$.

\subsection{Reflection spectrum}
\label{sec:ref}

Fig. \ref{fig:schem} illustrates the source irradiating a patch of
the disk that subtends a solid angle $d\Omega_d$ in the
\textit{source} rest frame and has a surface area $dA_d$ in the
reference frame of the \textit{disk patch}. Again using equation
\ref{eqn:fgen}, the specific flux crossing the surface of the disk
patch, in the restframe of the disk patch is
\begin{equation}
F_{d,in}(E_d,t_o') = \frac{C(t_o'-\tau_{sd})}{4\pi} g_{sd}^\Gamma f(E_d|\Gamma,g_{sd}E_{cut})
            \frac{d\Omega_{d}}{dA_{d}}.
	\label{eqn:Fdin}
\end{equation}
The irradiating flux is all re-processed into the reflection spectrum,
which is radiated an-isotropically from the disk upper surface
($\mu_e=\cos\delta_e \geq 0$; see Fig. \ref{fig:schem}). The emission
angle-dependent reflected specific intensity $I_{d,out}$ emergent from
the disk is related to the incident flux $F_{d,in}$ as
\begin{equation}
\int_0^\infty F_{d,in}(E_d,t_o') dE_d = 2\pi \int_0^1 \int_0^\infty I_{d,out}(E_d,t_o'|\mu_e) \mu_e dE_d d\mu_e.
\label{eqn:FinIout}
\end{equation}
As alluded to in the previous section, we use \textsc{xillver} or
\textsc{xillverCp} to calculate the reflected
specific intensity $\mathcal{R}(E|\mu_e)$ for an illuminating specific
flux $f(E|\Gamma,E_{cut})$ (we set the reflection fraction parameter
to $-1$, where the minus sign ensures that the \textsc{xillver} model
returns only the reflection spectrum rather than summing it with the
incident spectrum). The \textsc{xillver} models are normalized such
that
\begin{eqnarray}
	\frac{1}{2} \int_0^1 \int_0^\infty \mu_e
  \mathcal{R}(E|\mu_e,\Gamma,E_{cut},\log_{10}\xi) dE d\mu_e \nonumber \\
	= \int_0^\infty f(E|\Gamma,E_{cut},\log_{10}\xi) dE,
	\label{eqn:xilnorm}
\end{eqnarray}
where $\xi(r) = 4\pi F_x(r) / n_e(r)$ is the ionization parameter,
$F_x(r)$ is the $13.6$ eV to $13.6$ keV illuminating flux and $n_e(r)$
is the electron number density.

Inspection of equations (\ref{eqn:Fdin}), (\ref{eqn:FinIout}) and
(\ref{eqn:xilnorm}) shows
\begin{equation}
I_{d,out}(E_d,t_o'|\mu_e) = \frac{1}{2} \frac{C(t_o'-\tau_{sd})}{2\pi}
g_{sd}^\Gamma \frac{d\Omega_d}{dA_d} \frac{
  \mathcal{R}(E_d|\mu_e,g_{sd}E_{cut}) }{4\pi}.
\label{eqn:Idout}
\end{equation}
Once more exploiting the symmetry of the lamppost geometry, we can
consider the case whereby the disk patch is an annulus at $r$ with
width $dr$ to find
\begin{equation}
	\frac{d\Omega_d}{dA_d} = 2\pi \frac{|d\cos\delta/dr|}{dA_{\rm ring}/dr}.
\end{equation}
It is important to note that the angle $\delta$ in the equation is
defined in the source restframe, whereas the area $dA_{\rm ring}$
is defined in the restframe of the disk annulus. The radial coordinate
is defined in Boyer-Lindquist coordinates. We calculate $d\cos\delta/dr$ and $\tau_{sd}$ numerically
using \textsc{ynogk}. We calculate the area differential
analytically. In Boyer-Lindquist coordinates, the area of a disk
annulus with radial extent $dr$ is $d^2x = 2\pi \sqrt{g_{\phi\phi}
  g_{r r}} dr$. The area in the rest frame of the rotating annulus is
$dA_{\rm ring}=\gamma^\phi d^2x$, 
where $\gamma^\phi$ is the Lorentz factor of the annulus
(e.g. \citealt{Wilkins2012}). We present a derivation for
$\gamma^\phi$ in Appendix \ref{sec:area}, pointing out some very small
(largely inconsequential) errors in previous derivations
(\citealt{Bardeen1972,Dauser2013}).

According to the stationary observer, the disk patch centered at disk
coordinates $r$, $\phi$ is centered on coordinates on the observer
plane $\alpha$, $\beta$ and subtends a solid angle $d\Omega=d\alpha
d\beta/D^2$ (see Fig \ref{fig:schem}). Here $\alpha$ and $\beta$ are
the impact parameters at infinity. The specific flux seen from the
disk patch with coordinates $r$, $\phi$ by the stationary observer
viewing from an inclination angle $i$ (with $\mu=\cos i$) is therefore
$dR_o(E_o,t_o) = g_{do}^3 I_{d, out}(E_o/g_{do},t_o-\tau) d\alpha
d\beta /D^{2}$, giving
\begin{eqnarray}
	dR_o(E_o,t_o|\mu,r,\phi) = A[t_o-\tau(r,\phi)]
        g_{do}^3(r,\phi) \epsilon(r) \nonumber \\
\times \mathcal{R}[E_o/g_{do}(r,\phi)|\mu_e(r,\phi),g_{sd}(r) E_{cut},\log_{10}\xi(r)] d\alpha d\beta,
\label{eqn:dR}	
\end{eqnarray}
where
\begin{equation}
\epsilon(r) = \frac{g_{sd}^\Gamma(r)}{2}
\frac{|d\cos\delta/dr|}{dA_{\rm ring}/dr}
\label{eqn:eps}
\end{equation}
is the radial emissivity profile and
$\tau(r,\phi)=\tau_{sd}(r)+\tau_{do}(r,\phi)-\tau_{so}$. Note that for
equation \ref{eqn:dR} we have used equation
(\ref{eqn:liouville}), since $d\alpha d\beta/D^2$ is the solid angle
subtended by the disk patch according to the observer in the
observer's restframe. Also note that the cosine of the emission angle,
$\mu_e$, is a function of $r$ and $\phi$ because bending of rays leads
to photons with the same final trajectory having different initial
trajectories (\citealt{Garcia2014}). The total reflection spectrum
seen by the stationary observer is then calculated by integrating
equation (\ref{eqn:dR}) over all impact parameter values for
which the corresponding geodesic intercepts the disk at radii $r_{in}
< r < r_{out}$.

\subsection{Transfer function and cross-spectrum}
\label{sec:trans}

We can express the total reflected specific flux seen by the
$z=0$ observer as $R_o(E_o,t_o) = A(t_o) \otimes w_o(E_o,t_o)$, where
$\otimes$ denotes a convolution and
\begin{eqnarray}
w_o(E_o,t_o) = \int_\alpha \int_\beta \epsilon(r) g_{do}^3(r,\phi)
\delta(t_o-\tau(r,\phi) ) \nonumber \\
\times \mathcal{R}[E_o/g_{do}(r,\phi)|\mu_e(r,\phi),g_{sd}(r) E_{cut},\log_{10}\xi(r)] d\alpha d\beta.
\end{eqnarray}
is the \textit{impulse-response function}. In Fourier space the
convolution is a multiplication, so it is best to Fourier transform
the impulse response function to get the \textit{transfer function}
\begin{eqnarray}
W_o(E_o,\nu) = \int_\alpha \int_\beta \epsilon(r) g_{do}^3(r,\phi) {\rm
  e}^{i 2\pi \tau(r,\phi) \nu} \nonumber \\
\times \mathcal{R}[E_o/g_{do}(r,\phi)|\mu_e(r,\phi),g_{sd}(r) E_{cut},\log_{10}\xi(r)]
d\alpha d\beta.
\label{eqn:transo}
\end{eqnarray}
Setting $\nu=0$ (The DC component, standing for direct current), gives
the time-averaged spectrum. Generalising to an observer at a
cosmological distance gives $R(E,\nu) = A(\nu) W(E,\nu)$, where
\begin{eqnarray}
W(E,\nu) = (1+z)^{-3} \int_\alpha \int_\beta \epsilon(r) g_{do}^3(r,\phi) {\rm
  e}^{i 2\pi (1+z) \tau(r,\phi) \nu} \nonumber \\
\times \mathcal{R}[E_o/g_{do}(r,\phi)|\mu_e(r,\phi),g_{sd}(r) E_{cut},\log_{10}\xi(r)]
d\alpha d\beta,
\label{eqn:trans}
\end{eqnarray}
and we note that $A(\nu)$ is in general complex. We trace rays defined
by given impact parameter values backwards from the observer
plane towards the black hole along the relevant null geodesic in the
Kerr metric (again calculated using \textsc{ynogk}). This operation
automatically accounts for lensing of rays travelling from the disk to
the observer. We consider a
$400\times 400$ grid of impact parameters with
$\sqrt{\alpha^2+\beta^2} \leq 300~R_g$. We additionally consider a
larger grid with $300~R_g < \sqrt{\alpha^2+\beta^2} \leq r_{out}$ for
which we calculate geodesics in the Minkowski metric. For rays that
cross the disk mid-plane, we calculate the $r$, $\phi$ and $t$
coordinates at the crossing point. We stop following rays after the
first time they cross the mid-plane, therefore ignoring ghost images,
which are likely blocked in reality by material in the vicinity of the
hole. We quote the formulae for the blueshift factors and angles in
Appendix \ref{sec:gandmu}. We  also include a `boosting factor' to
account for the likelihood that our assumption of an isotropically
radiating source is inappropriate. We specify the factor $1/\mathcal{B}$ as a model
parameter, such that $1/\mathcal{B}<1$ roughly corresponds to the
source being beamed towards us and away from the disk.

From the transfer function, we can calculate the energy-dependent
cross-spectrum. This is a series of cross-spectra between the flux at
each energy and that in a common reference band, $G(E,\nu) = S(E,\nu)
F_r^*(\nu)$. In our model, this is given by
\begin{equation}
G(E,\nu) =
\alpha(\nu) {\rm e}^{i \phi_A(\nu)} \left[ F_{A=1}(E) +
  \frac{W(E,\nu)}{\mathcal{B}} \right],
\label{eqn:G}
\end{equation}
where $\alpha(\nu)$ and $\phi_A(\nu)$ are model parameters for each
frequency $\nu$, and $F_{A=1}(E)$ is given by equation (\ref{eqn:F})
with $A=1$. We could equally see equation \ref{eqn:G} as a formula for
the \textit{complex-covariance} (\citealt{Mastroserio2018}), which is
simply the cross-spectrum divided through by the amplitude of the
reference band, $S(E,\nu) F_r(\nu) / |F_r(\nu)|$. The only adjustment
would be a slight change in the, already fairly arbitrary, meaning of
the normalisation parameter $\alpha(\nu)$. Finally, line-of-sight
absorption is accounted for using the multiplicative \textsc{xspec}
model, \textsc{tbabs} \citep{Wilms2000}, such that the transmitted
cross-spectrum is $G_{abs}(E,\nu) = \textsc{tbabs}(E) \times
G(E,\nu)$. For a given
frequency range, our model calculates this transmitted cross-spectrum
and outputs, as a function of energy, the real part of this, the
imaginary part, the modulus (energy-dependent variability amplitude)
or the time lag ($t_{\rm lag}(E,\nu)={\rm
  arg}\{G(E,\nu)\}/[2\pi\nu]$), depending on the user-defined mode.

As discussed in \cite{Mastroserio2018}, if the reference band is the
sum of energy channels ranging from $I_{min}$ to $I_{max}$ that are
all well calibrated for the instrument we are observing with, the
parameter $\phi_A(\nu)$ need not be a free parameter, and can instead
be expressed as
\begin{equation}
\tan\phi_A(\nu) = \frac{ -\sum_{J=I_{min}}^{I_{max}} \Im[W(J,\nu)]/\mathcal{B}  }
{ \sum_{I=I_{min}}^{I_{max}} F_{A=1}(I) + \Re[W(I,\nu)]/\mathcal{B}  }.
\label{eqn:phiA}
\end{equation}
Here, $F_{A=1}(I)$ and $W(I,\nu)$ are calculated by convolving
$\textsc{tbabs}(E)\times F_{A=1} (E)$ and $\textsc{tbabs}(E)\times
W(E,\nu)$ respectively with the instrument response (see
\citealt{Mastroserio2018} for details). Our model incorporates both a
mode in which $\phi_A(\nu)$ is a free parameter and a mode in which
the instrument response is read in and $\phi_A(\nu)$ is calculated
self-consistently.

\subsection{The reflection kernel}
\label{sec:kernel}

Much of the computational expense of evaluating equation
(\ref{eqn:transo}) can be saved by representing it as a convolution
with a kernel. The easiest case to calculate is if we assume that the
shape of the rest frame reflection spectrum depends on neither $r$ nor
$\phi$. This can be done by assuming that the cut-off energy
seen by each disk patch is $E_{cut}$ rather than $g_{sd}(r)E_{cut}$,
that $\delta_e(r,\phi)=i$ and that the disk ionization parameter is
independent of radius. Working with $\log E$ rather than $E$, allows
the transfer function to be represented as
\begin{equation}
W(\log E, \nu) = \int_0^\infty \mathcal{R}(\log E') W_{\delta}(\log(E/E'),\nu) {\rm d}\log E',
\label{eqn:trans2}
\end{equation}
where
\begin{eqnarray}
W_{\delta}(\log E,\nu) = (1+z)^{-3} \int_{\alpha,\beta} \epsilon(r)
  g_{do}^3(r,\phi) {\rm e}^{i 2\pi (1+z) \tau(r,\phi) \nu} \nonumber \\
\times \delta\left[\log E-\log\left( \frac{g_{do}(r,\phi)}{1+z} \right)\right] {\rm d}\alpha {\rm d}\beta
\label{eqn:kernel}
\end{eqnarray}
is the kernel of the transfer function. It is clear that the kernel is
simply the transfer function for a $\delta-$function rest frame
reflection spectrum centered at $1$ keV.  Equation (\ref{eqn:trans2})
can be recognised as a convolution in $\log E$ space, and can thus be
written
\begin{equation}
W(\log E, \nu) = \mathcal{R}(\log E) \otimes_{\log E} W_{\delta}(\log
E,\nu).
\label{eqn:WlogE}
\end{equation}
We compute the convolution using the convolution theorem (i.e. Fourier
transforming both, multiplying and finally inverse transforming),
which allows us to exploit the large gain in speed afforded by using
the fast Fourier transform (FFT) algorithm.

In the more general case, we can quantise $\mu_e$, $g_{sd}$ and
$\log_{10}\xi$ by defining a number of discrete bins for each and
writing the transfer function as
\begin{eqnarray}
W(\log E, \nu) = \sum_{j} \sum_{k} \sum_{n} \mathcal{R}[\log
E|\mu_e(j),g_{sd}(k),\log_{10}\xi(n)] \nonumber \\
\otimes_{\log E} 
W_{\delta}(\log E,\nu|j,k,n),
\label{eqn:Wfancy}
\end{eqnarray}
where $W_{\delta}(\log E,\nu|j,k,n)$ is given by equation
\ref{eqn:kernel}, except the integrand is only non-zero for disk
patches with $\mu_e(r,\phi)$, $g_{sd}(r)$ and $\log_{10}\xi(r)$ in
the range specified by the indices $j$, $k$ and $n$. Computing
equation \ref{eqn:Wfancy} therefore requires a convolution for each
permutation of $j$, $k$ and $n$. Use of the FFT algorithm prevents the
computation of so many convolutions from becoming prohibitively
expensive. We will explore the effect of changing the number of
convolutions on the accuracy of the model in section
\ref{sec:details}.

\begin{figure*}
  \begin{centering}
	\includegraphics[angle=0,width=12.5 cm,trim= 2.0cm 1.2cm 2.5cm
        8.5cm,clip=true]{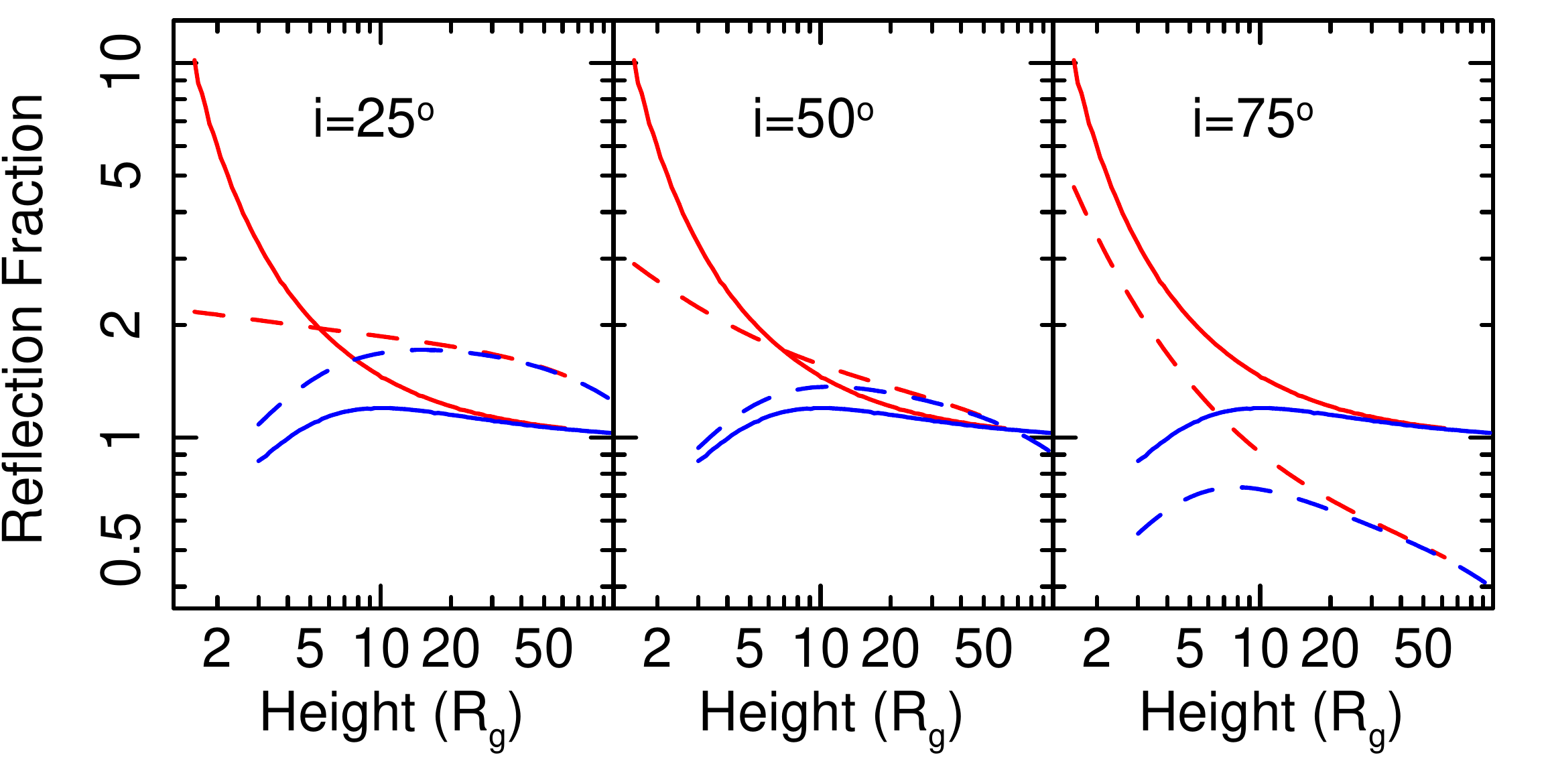} \hspace{-2mm}
	\includegraphics[angle=0,width=4.5 cm,trim=1.5cm 1.5cm
        9.5cm 11.0cm,clip=true]{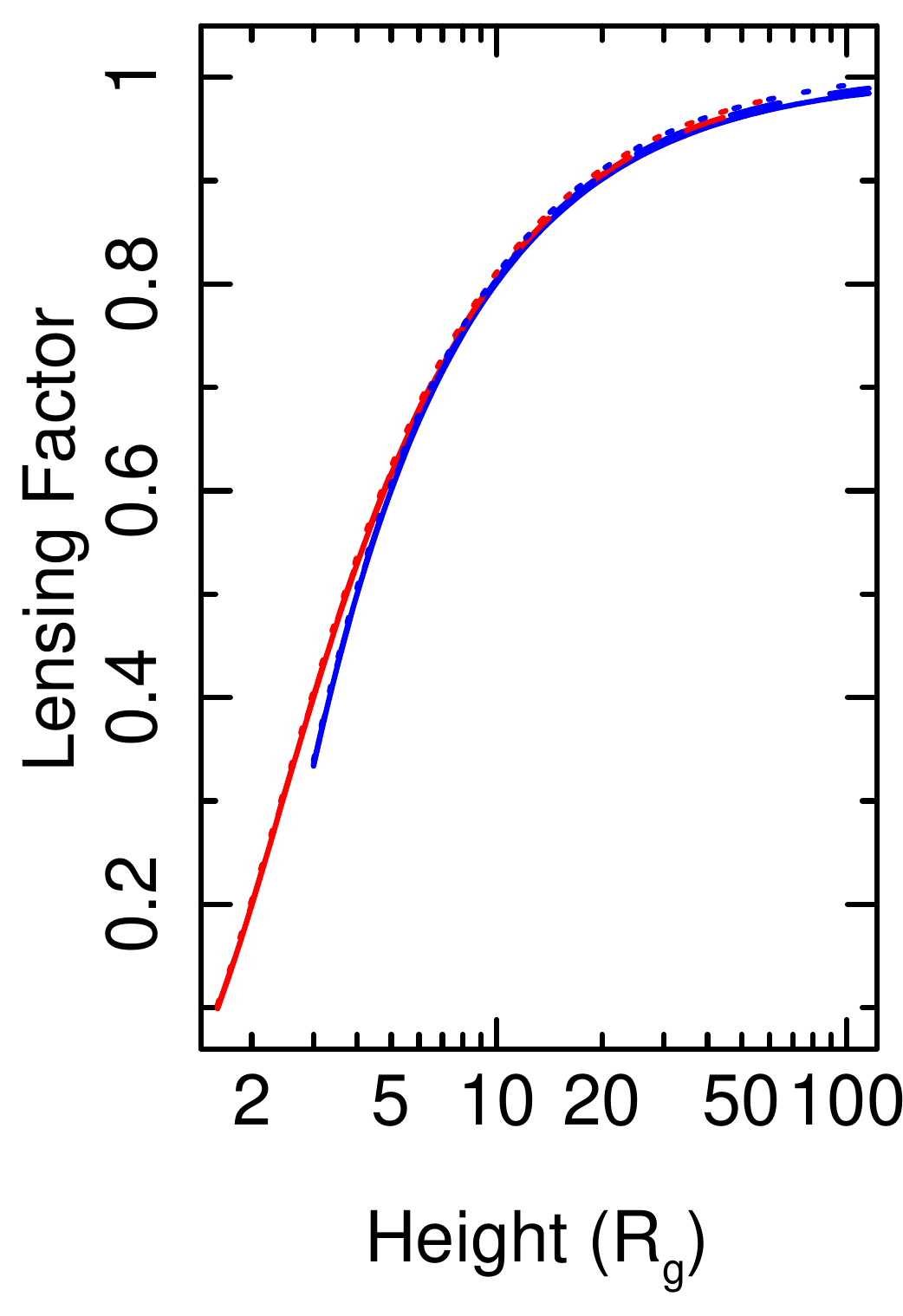}
  \end{centering}
\vspace{-0mm}
 \caption{\textit{Left:} Reflection fraction plotted against source
   height for three inclination angles, with the disk extending down
   to the ISCO for black hole spin $a=0.998$ (red) and $a=0$
   (blue). The solid lines are `system' reflection fraction as defined
   by \citet{Dauser2016} (see our equation \ref{eqn:frrelxill}) and
   the dashed lines are the observer's reflection fraction as defined
   by our equation \ref{eqn:frobs}. The system reflection fraction
   does not depend on inclination angle whereas the observer's
   reflection fraction does, even in the Newtonian
   limit. \textit{Right:} Lensing factor, $\ell$, plotted against
   source height for $a=0.998$ (red) and $a=0$ (blue) with
   $i=25^\circ$ (solid), $i=50^\circ$ (dashed) and $i=75^\circ$
   (dotted). We see that $\ell$ depends very strongly on $h$, but only
   weakly on $a$ and $i$.}
 \label{fig:refl_frac}
\end{figure*}

\begin{table}
\begin{center}
\begin{tabular}{|p{1.2cm}|p{1.5cm}|p{3cm}|p{1.0cm}|} 
\hline
Parameter  & Units & Description & Default value \\
\hline
\hline
$h$ & $R_g$ (+ve) or $R_h$ (-ve) & Source height & $6$ \\
\hline
$a$   &  & Dimensionless spin parameter & $0.9$ \\
\hline
$i$ & Degrees & Inclination angle & $30$ \\
\hline
$r_{in}$ & $R_g$ (+ve) or ISCO (-ve) & Disk inner
                                                      radius & $-1$ \\
\hline
$r_{out}$ & $R_g$ & Disk outer radius & $20000$ \\
\hline
$z$ & & Cosmological redshift & $0$ \\
\hline
$\Gamma$ & & Photon index & $2$ \\
\hline
$\log_{10}\xi$ & $\xi$ has units of erg~cm/s & Ionization parameter or peak
                                    value of ionization parameter & $3$ or $3.75$\\
\hline
$A_{Fe}$ & Solar & Relative iron abundance & $1$ \\
\hline
$(E_{cut})_{obs}$ & keV & Observed high energy cut-off & $300$ \\
\hline
$(kT_{e})_{obs}$ & keV & Observed electron temperature & $300$ \\
\hline
$N_h$ & $10^{22}{\rm cm}^{-2}$ & Hydrogen column density of material
                                 in the line-of-sight (\textsc{tbabs}) & $0$ \\
\hline
$1/\mathcal{B}$ & & Boosting factor to adjust the
                                     reflection fraction from lamppost
                                        value & $1$ \\
\hline
$M$ & $M_\odot$ & Black hole mass & $4.6\times 10^{7}$ \\
\hline
$\phi_A$ & Radians & Phase norm - can be
                                     self-consistently calculated & $0$ \\
\hline
$\nu_{min}$ & Hz & Minimum frequency transfer function
                                        is averaged over & $1 \times 10^{-5}$ or $0$\\
\hline
$\nu_{max}$ & Hz & Maximum frequency transfer function
                                        is averaged over & $2 \times 10^{-5}$ or $0$\\
\hline
\texttt{ReIm} &   &  Sets output & $1-6$\\
\hline
\end{tabular}
\end{center}
\caption{Model parameters for \textsc{reltrans} and
  \textsc{reltransCp}. Source height and disk inner radius can be
  expressed in horizons and ISCOs respectively in order to avoid
  unphysical parameter combinations during exploration of parameters
  space. The chosen value of mass corresponds to NGC 4151. The model
  calculates the energy dependent cross-spectrum averaged over the
  frequency range $\nu_{min}$ to $\nu_{max}$. The parameter
  \texttt{ReIm} sets the model output. The options are: 1) real part,
  2) imaginary part, 3) modulus (i.e. the absolute variability
  amplitude), 4) time lag (the argument divided by $2\pi \nu$, where
  $\nu=[\nu_{min}+\nu_{max}]/2$), 5) modulus of the folded
  cross-spectrum, and 6) the time-lag for the folded
  cross-spectrum. If either $\nu_{min}$ or $\nu_{max}$ are set to
  zero, the time-averaged spectrum is returned.}
\label{tab:par}
\end{table}

\subsection{Reflection fraction}
\label{sec:reflfrac}

It is useful to define a reflection fraction that captures the ratio
between reflected and direct components in the observed spectrum,
specifically isolating geometric considerations from radiative
transfer considerations. Here, we discuss two definitions of the
reflection fraction: the \textit{system reflection fraction}, which
depends only on the geometry of the system and is independent of the
observer, and the \textit{observer's reflection fraction}.

The system reflection fraction, already used by the model
\textsc{relxilllp} (\citealt{Dauser2013}), is
\begin{equation}
(f_R)_{sys} = \frac{ \cos\delta_{in} - \cos\delta_{out} } { 1 +
  \cos\delta_{out} },
\label{eqn:frrelxill}
\end{equation}
where $\delta_{in}$ and $\delta_{out}$ are respectively the values of
the angle $\delta$ for geodesics from the source that hit the inner
and outer radii of the disk \citep{Dauser2014}. This definition gives the
number of photons that hit the disk divided by the number that reach
infinity in the hemisphere above the disk mid-plane. In the case of
Newtonian gravity, $(f_R)_{sys}$ would reach a maximum of unity for a
disk extending from $r_{in}=0$ to $r_{out}=\infty$. In full GR
however, $(f_R)_{sys}$ can be much larger due to focusing of photons
onto the inner regions of the disk (\citealt{Dauser2016}).

Although the above definition is conveniently simple, it does not
fully capture the relative flux of the direct and reflected spectra as
seen by a given observer. We therefore additionally define an
observer's reflection fraction. In order to exclude the radiative
transfer calculation, we define a reflection spectrum for the case in
which the disk re-emits the incident spectrum isotropically. In this
case, we can define the reflection fraction as the \textit{observed bolometric
  reflected flux divided by the directly observed bolometric
  flux}. Note that both of these fluxes are considered to be measured
in the \textit{observer's frame}. This means that the
specific flux re-radiated from a disk patch is $F_{d,out}(E_d) =
F_{d,in}(E_d)$ (input spectrum preserved) and the specific intensity
re-radiated from the disk patch is $I_{d,out}(E_d) =  F_{d,in}(E_d)
/ \pi$ (isotropic re-radiation). This gives
\begin{eqnarray}
(f_R)_{obs} &=& \frac{2}{\ell g_{so}} \int_{\alpha,\beta} g_{do}^3(r,\phi) g_{sd}(r)
\frac{|d\cos\delta/dr|}{dA_{\rm ring}/dr} d\alpha d\beta \nonumber \\
&=& \frac{4}{\ell g_{so}} \int_{\alpha,\beta} g_{do}^3(r,\phi) g_{sd}^{1-\Gamma}(r)
\epsilon(r) d\alpha d\beta
\label{eqn:frobs}
\end{eqnarray}

Applying the earlier experiment of taking the simple limiting case of
an infinite slab in Newtonian gravity to equation (\ref{eqn:frobs})
gives $(f_R)_{obs}=2\cos i$.
Averaging over all $\cos i$ in the
hemisphere above the disk mid-plane, we find $\langle (f_R)_{obs}
\rangle = 1$; i.e. source photons are either radiated into the upper
hemisphere to be observed directly, or into the lower hemisphere to be
observed as reflection. The angular dependence, even when isotropic
radiation is assumed, results from the source being a sphere, whereas
the disk is a slab. This definition of the reflection fraction is
similar, but not identical, to the `reflection strength' defined by
\cite{Dauser2016}. In our model, we calculate both reflection
fractions and print them to screen if the verbose level is set
suitably high by an environment variable.

Fig \ref{fig:refl_frac} (left) shows our two definitions of reflection
fraction plotted against source height for a number of parameter
combinations. We see that the solid lines representing $(f_r)_{sys}$,
which agree exactly with Fig 3 of \cite{Dauser2016}, do not depend on
viewer inclination. The dashed lines representing 
$(f_r)_{obs}$ do depend on viewer position. The right panel shows the
contribution of the lensing factor. This is strongly dependent on
source height, but only weakly dependent on spin and inclination.

\section{Model properties}
\label{sec:results}

The model parameters of \textsc{reltrans} and \textsc{reltransCp} are
listed in Table \ref{tab:par}. We also define a number of environment
variables, listed in Appendix \ref{sec:env}, that are used to switch
between different modes and control resolution. Each environment
variable has a sensible default value, such that the model is
user friendly for the beginner and flexible for the advanced
user. In this section, we explore the model properties and describe
the listed parameters and environment variables. For the sake of 
intuition we plot time lags and variability amplitudes, even though it
is statistically favourable when fitting to data to consider real and
imaginary parts of the cross-spectrum
(\citealt{Ingram2016,Ingram2017,Rapisarda2017,Mastroserio2018}).

\subsection{Emissivity profiles}
\label{sec:emissivity}

\begin{figure}
  \begin{centering}
	\includegraphics[angle=0,width=\columnwidth,trim= 1.8cm 1.5cm 4.2cm 3.0cm,clip=true]{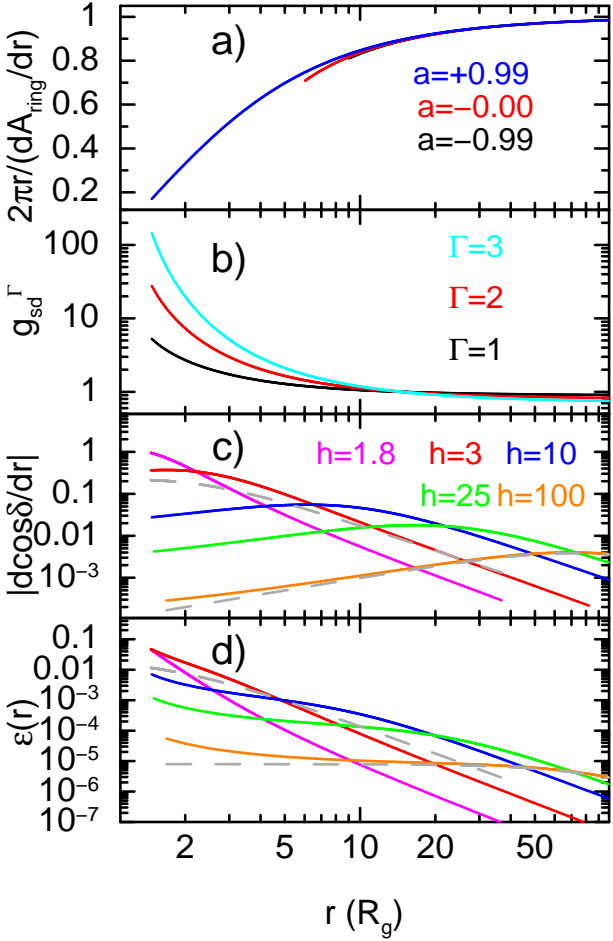}
  \end{centering}
\vspace{-5mm}
 \caption{Contributions to the radial emissivity profile, designed for
   comparison to Fig 2 in Dauser et al (2013). \textit{a:} Ratio of
   the Newtonian radial derivative to the fully relativistic version for spin
 as labelled. \textit{b:} Contribution to the emissivity of the
 blueshift factor for $h=10$ and $a=0.99$. \textit{c:}
 Radial derivative of $\cos\delta$ for $a=0.99$ and $h$ as
 labelled. The grey dashed lines are the Newtonian equivalent for
 $h=1.8$ (to be compared with the solid magenta line) and $h=100$ (to
 be compared with the solid orange line). As expected, this is a better
 approximation for larger source heights. \textit{d:} Emissivity
 profile for $\Gamma=2$, $a=0.99$ and $h$ as labelled in panel
 c. Again, the grey dashed lines are the Newtonian equivalent
 [$\epsilon(r) \propto (h^2+r^2)^{-3/2}$] for $h=100$ and $h=1.8$.}
 \label{fig:emissivity}
\end{figure}

Fig \ref{fig:emissivity} shows the lamppost model emissivity profile
and some contributing factors for a range of parameter
combinations. Panel (a) shows the ratio of the area derivative in the
Newtonian case to the relativistic case for three different values of
spin. The difference between the three spin values results entirely
from the Lorentz factor of the rotating disk element,
$\gamma^\phi$. This plot is very similar, but not identical, to the
corresponding plot in \cite{Dauser2013} (top panel in their Fig
2). The discrepancy results from small (inconsequential as it turns
out) mistakes in the expressions for $\gamma^\phi$ in
\cite{Bardeen1972} (equation 13.12a) and \cite{Dauser2013} 
(equation 10). The two expressions are identical \textit{except} the
latter reference drops all $\pm$ and $\mp$ signs, meaning that they
agree for prograde spin but differ slightly for retrograde spin. Upon
further investigation, detailed in Appendix \ref{sec:area}, we found a
very subtle mistake in equation (13.12a) of \cite{Bardeen1972}, which
is again very small and only relevant for retrograde spin. We find
that, for $a=-0.99$ (which maximises the magnitude of the error), the
\cite{Dauser2013} version actually gives a closer answer to our new
expression than the \cite{Bardeen1972} version, although all three are
very similar.

Panel (b) shows the contribution of the blueshift factor for three
different values of $\Gamma$, illustrating that a steeper spectrum
leads to a steeper emissivity profile. Panels (c) and (d) show
respectively the radial derivative of the cosine of the angle $\delta$
and the overall emissivity profile for various parameter
combinations. The grey dashed lines represent the Newtonian
approximations [$|d\cos\delta/dr|= h r (h^2+r^2)^{-3/2}$ and
$\epsilon(r) = h (h^2+r^2)^{-3/2} / (4\pi)$] for $h=1.8~R_g$ (to be
compared with the magenta lines) and $h=100~R_g$ (to be compared with
the orange lines). We see that, as expected, the full GR solution
diverges dramatically from the Newtonian approximation for low source
heights. Our emissivity profiles agree with those presented in
\cite{Dauser2013}.

\subsection{Time-averaged spectrum}
\label{sec:speccomp}

\begin{figure}
  \begin{centering}
	\includegraphics[angle=0,width=\columnwidth,trim=1.5cm 1.5cm
        2.5cm 11.0cm,clip=true]{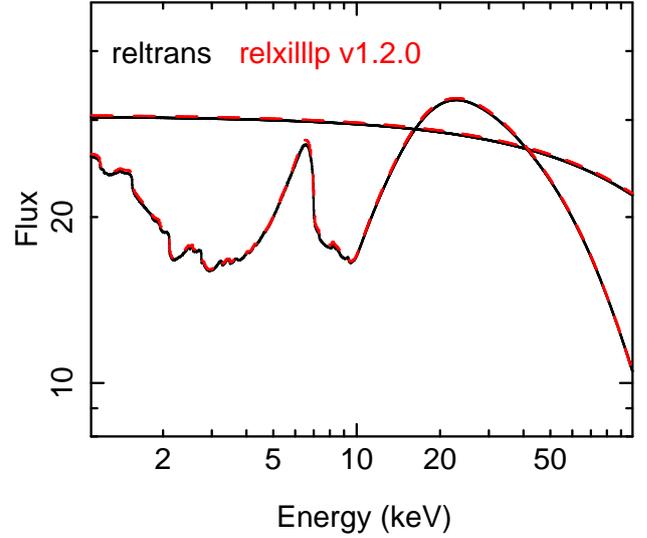}
  \end{centering}
\vspace{-5mm}
\caption{Time-averaged direct and reflected spectrum for
  \textsc{reltrans} (black), and the most recent version of
  \textsc{relxilllp} (red, dashed). We see good agreement between the
  two models.}
 \label{fig:benchmark}
\end{figure}

Fig \ref{fig:benchmark} shows the direct and reflected
components of the time-averaged spectrum calculated by
\textsc{reltrans} (black) and the most recent version of
\textsc{relxilllp} at the time of writing (red, dashed). We use the
default parameters listed in Table \ref{tab:par}, except we set
$r_{out}=400~R_g$ for ease of comparison with
\textsc{relxilllp}. \textsc{relxilllp} accounts for the dependence of
emission angle and disk rest frame cut-off energy on disk coordinates,
but assumes a single ionization parameter. For the purposes of
comparison in this plot, we therefore follow suit (although see
sections \ref{sec:mue} and \ref{sec:ion} for further discussion on
these dependencies), and use the default number of zones for both 
$\mu_e$ and $E_{cut}$. For all models, the relative normalisation of
reflected and direct components is calculated self-consistently, rather than
set as a model parameter. We see that \textsc{relxilllp} agrees very
well with our model\footnote{Model versions of \textsc{relxillp} prior
  to v1.2.0 have a smaller relative normalization of the  reflection
  spectrum. This comes partly from an extra factor of $0.5\cos i$ that
  is applied to the \textsc{xillver} spectrum before being convolved
  with the smearing kernel in the older versions}. Besides
benchmarking against \textsc{relxilllp}, we have also throughly tested
our code by comparing it with outputs calculated using brute-force
calculation of the transfer function (i.e. without using the kernal
convolution).

\subsection{Rest frame assumptions}
\label{sec:details}

In this section, we explore the impact of accounting for the
coordinate dependence of the emission angle and high energy cut-off
(\ref{sec:mue}) and ionization parameter (\ref{sec:ion}), before
comparing \textsc{reltrans} to \textsc{reltransCp} (\ref{sec:Cp}).

\subsubsection{Emission angle and cut-off energy}
\label{sec:mue}

\begin{figure}
  \begin{centering}
	\includegraphics[angle=0,width=\columnwidth,trim=1.5cm 1.5cm
        2.5cm 11.0cm,clip=true]{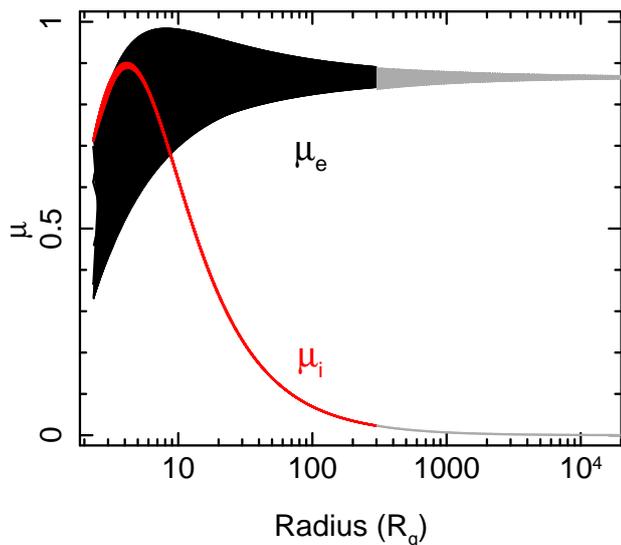}
  \end{centering}
\vspace{-8mm}
 \caption{Cosine of the emission angle (black) and incidence angle
   (red) as a function of radius for the default parameters ($i=30^\circ$). We see
   that the emission angle depends on azimuth as well as radius,
   whereas the incidence angle is a monotonic function of radius. The
   grey points at $r\gtrsim 400~R_g$ are computed assuming that rays
   travel in straight lines. The smooth joins from the full GR
   treatment used for $r \lesssim 400~R_g$ demonstrate that this is a
   reasonable assumption.}
 \label{fig:mu}
\end{figure}



\begin{figure*}
  \begin{centering}
	\includegraphics[angle=0,width=\columnwidth,trim= 1.7cm 1.5cm 1.8cm 3.0cm,clip=true]{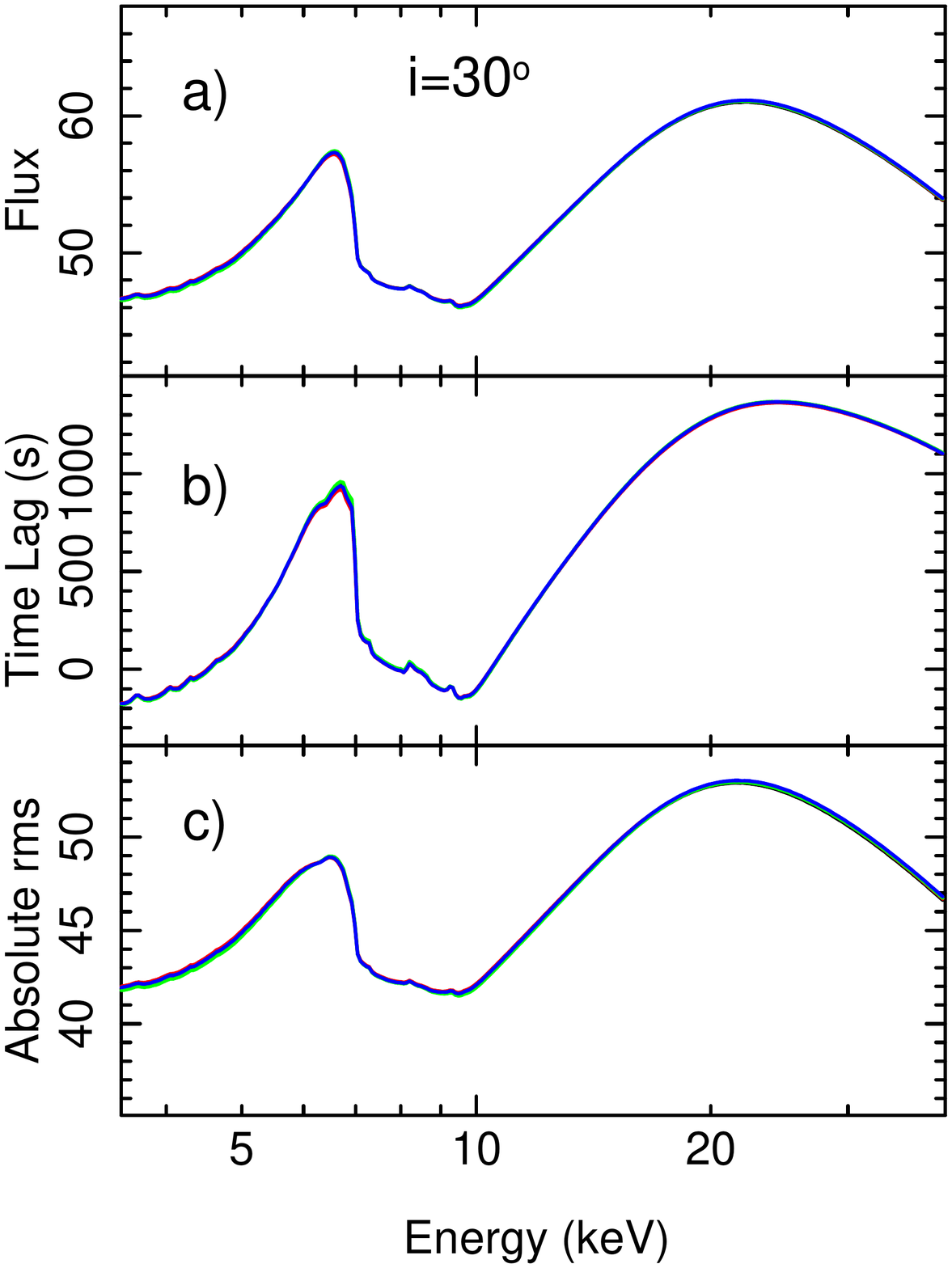} ~~~~
	\includegraphics[angle=0,width=\columnwidth,trim= 1.7cm 1.5cm 1.8cm 3.0cm,clip=true]{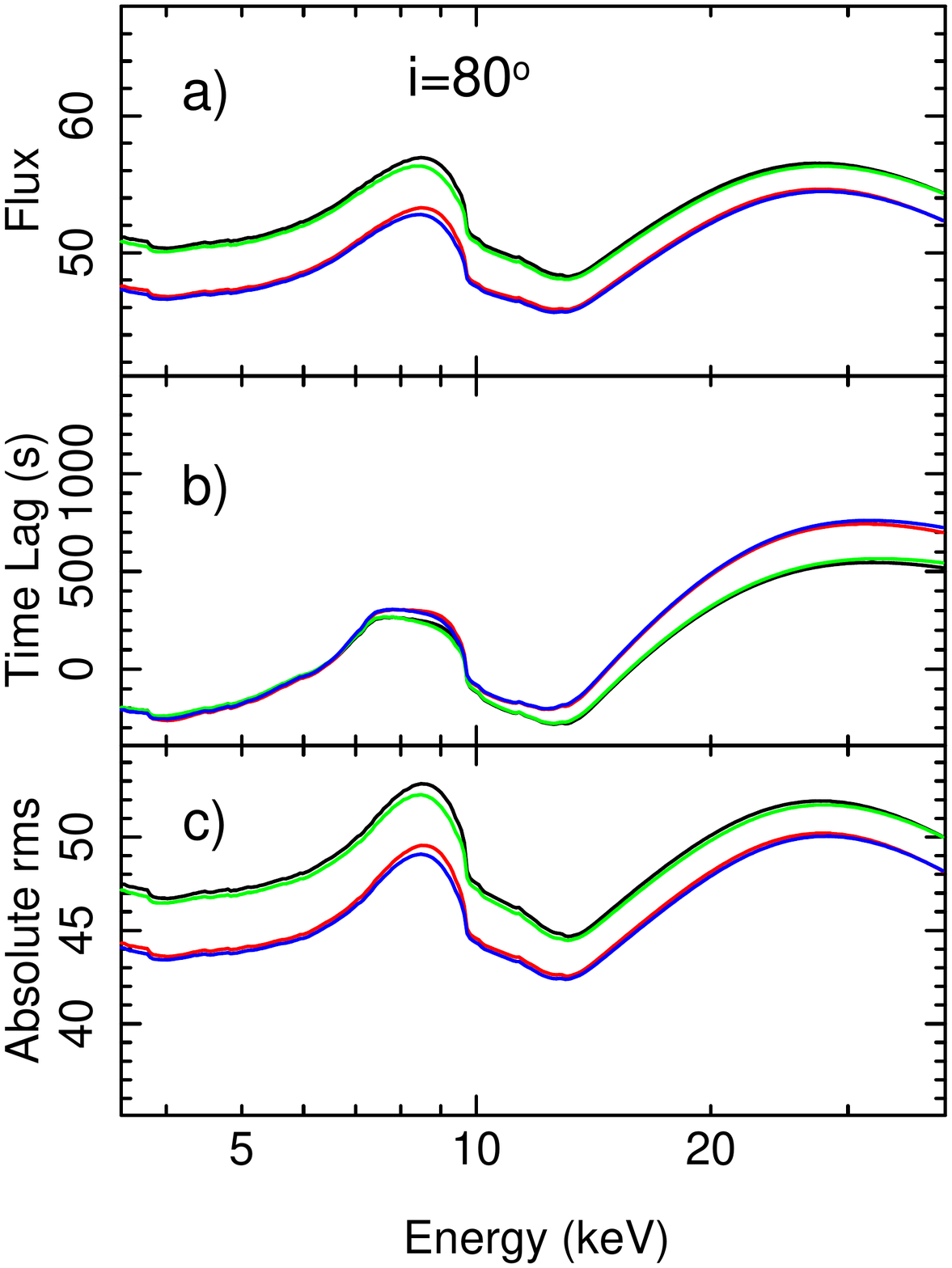}        
  \end{centering}
\vspace{-5mm}
 \caption{\textsc{reltrans} time-averaged spectrum (a), $1-2 \times
   10^{-5}$ Hz time lags (b) and absolute variability amplitude (c)
   calculated for the default parameters. Left and right hand panels
   correspond to inclination angles of $30^\circ$ and $80^\circ$
   respectively. For the black lines, we set the emission angle 
   $\delta_e$ equal to the inclination angle $i$ and ignore the radial
   dependence of $E_{cut}$ as measured in the disk restframe. For the
   other lines, we account only for the disk coordinate dependence of
   $\mu_e$ (red), only for the $E_{cut}$ dependence (green), and for
   both (blue). A single ionization parameter is used. For the lags,
   $\phi_A$ is calculated for a $2-10$ keV \xmm~EPIC pn reference band
   flux.}
 \label{fig:mueandg}
\end{figure*}

Fig \ref{fig:mu} shows the radial dependence of the cosine of the
emission angle, $\mu_e$, for the
default parameters, with the spread being for different disk
azimuths. As expected, $\mu_e \approx \cos(i)$
for very large disk radii, but covers an enormous range for
smaller disk radii. The \textsc{relxill} family of models for the
time-averaged spectrum (\citealt{Garcia2014}) account for this disk
coordinate dependence of the emission angle, and now also for the
radial dependence of apparent $E_{cut}$ observed in the disk rest
frame. Here, we investigate both effects in the context of the timing
properties. Fig \ref{fig:mueandg} shows the time-averaged spectrum
(a), time lags (b) and variability amplitude (c) calculated for
$i=30^\circ$ (left) and $i=80^\circ$ (right). The different lines
account for neither effect (black), only emission angle (red), only
cut-off energy (green) and both effects (blue). For the purposes of
the variability amplitude calculation, we simply set $\alpha(\nu)=1$
(this in itself is unphysical, corresponding to 100\% fractional
variability, but as an arbitrary constant it does not have any bearing
on our analysis). We see that the $E_{cut}$ effect is always subtle,
but the emission angle effect can become very large for high
inclination angles (consistent with what \citealt{Garcia2014} found for the
time-averaged spectrum). These figures use 10 zones to account for
both effects, which we find to be comfortably enough to reach
convergence for all trialled parameter combinations. Since acceptable
accuracy can be achieved by using as few as 5 zones for both
quantities, we set 5 as the default value for the environment
variables \texttt{MU\_ZONES} and \texttt{ECUT\_ZONES} (see Table
\ref{tab:env}). The user can adjust these values to test for
convergence.

\subsubsection{Ionization profile and incidence angle}
\label{sec:ion}

\begin{figure}
  \begin{centering}
	\includegraphics[angle=0,width=\columnwidth,trim=1.5cm 1.5cm
        2.5cm 11.0cm,clip=true]{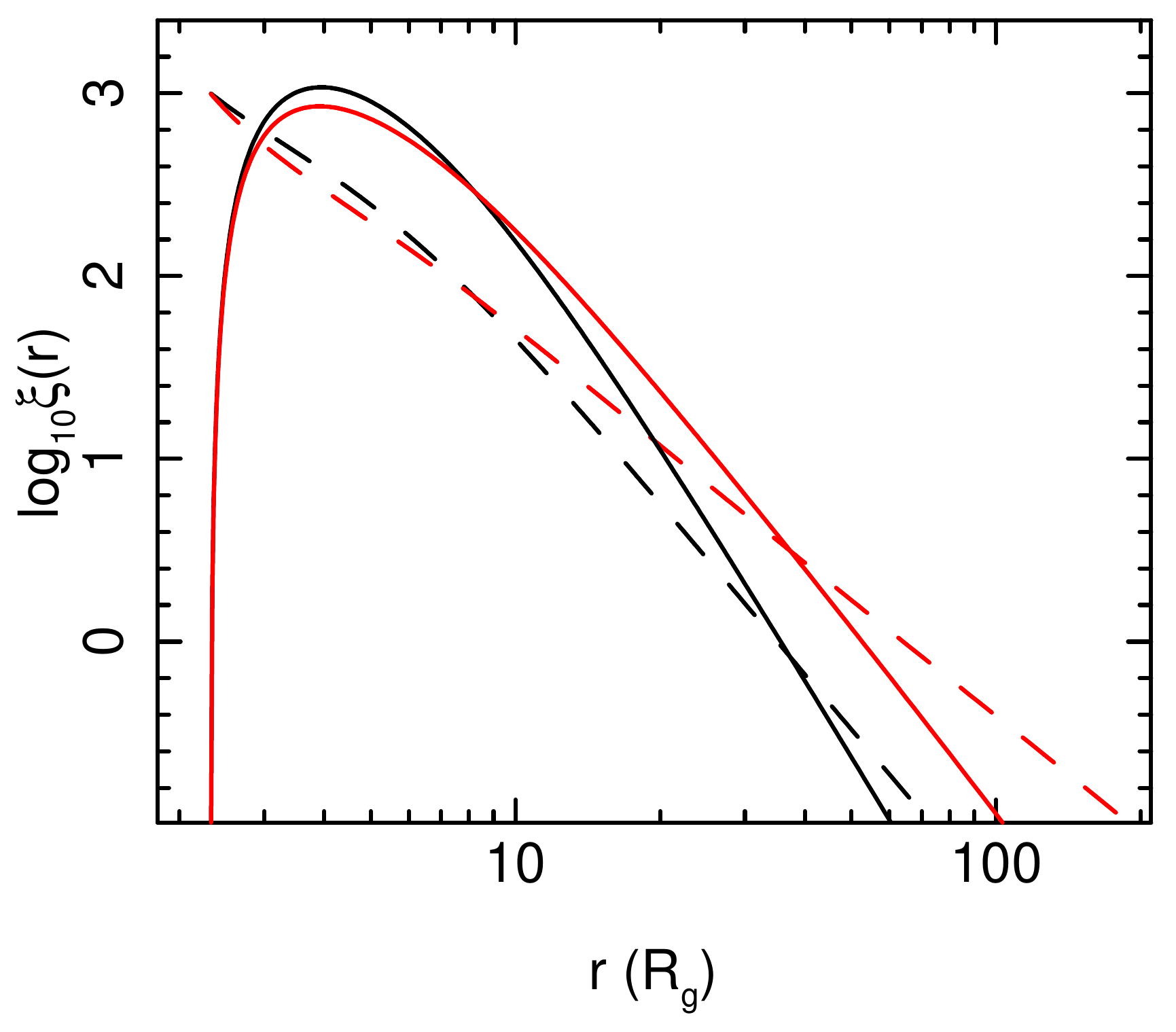}
  \end{centering}
\vspace{-5mm}
 \caption{The black lines are radial ionization profiles calculated
   assuming a zone A density profile (solid) and constant density
   (dashed). The red lines are effective ionization profiles, which
   have been adjusted to account for the radial dependence of the
   incidence angle.}
 \label{fig:xivsr}
\end{figure}

The ionization parameter is proportional to the $13.6$ eV to $13.6$
keV illuminating flux,
$F_x(r) \propto g_{sd}^{2-\Gamma}(r) \epsilon(r)$, divided by the disk
electron density $n_e(r)$. Whereas the flux is known exactly in the
lamppost model, $n_e(r)$ is more uncertain. \cite{Shakura1973} define
equations for 3 disk zones, where zone A is the innermost region, in
which pressure is dominated by radiation and the opacity is dominated
by electron scattering. Since the emissivity is dominated by the
inner regions, we first investigate the zone A density profile
(equation 2.11), $n_e(r) \propto \alpha^{-1} r^{3/2} [ 1 - ( r_{in}/r
)^{1/2} ]^{-2}$, where $\alpha$ is the viscosity parameter [not to be
confused with our normalisation parameter $\alpha(\nu)$]. The density
profile is very different for the other zones at larger radii, but for
these radii $F_x$ is small and so the predicted ionization is small
regardless of the assumed density profile. The black solid line in Fig
\ref{fig:xivsr} shows the resulting ionization profile for the
default parameters. For simplicity, we have taken the viscosity
parameter to be a constant, but we stress there is no \textit{a
  priori} reason to assume this. There are other reasons to suspect an
alternative density profile. For instance, the stress-free inner
boundary condition may not be appropriate for a truncated disk, or
there may be no zone A present when the accretion rate is a small
fraction of the Eddington limit. We therefore additionally explore the
simplest possible case of constant density (following
\citealt{Svoboda2012}). The resulting ionization profile is plotted in
Fig \ref{fig:xivsr} (black dashed line).

We normalise the ionization profile by specifying as a model parameter
the peak ionization value, $\log_{10}\xi_{max}$. For the constant
density model, the peak simply occurs at the disk inner radius. For
the zone A density profile, we use $r_{peak} = (11/9)^2 r_{in}$, which
is only exact for $\epsilon(r) \propto r^{-3}$, but numerical
calculation of the exact $r_{peak}$ would be fairly expensive for no
real gain.

\begin{figure}
  \begin{centering}
	\includegraphics[angle=0,width=\columnwidth,trim= 1.7cm 1.5cm 1.8cm 3.0cm,clip=true]{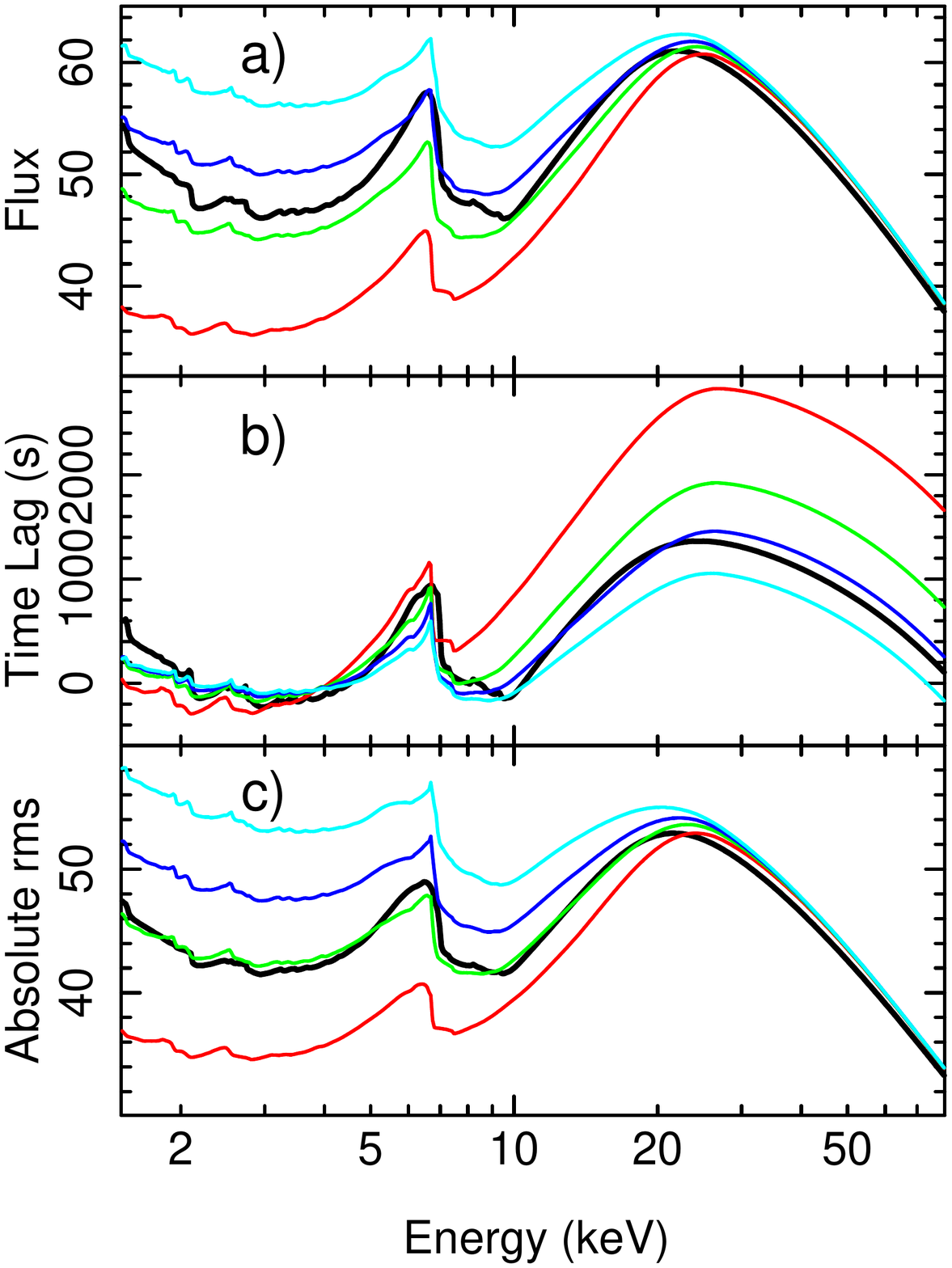}
  \end{centering}
\vspace{-8mm}
 \caption{\textsc{reltrans} time-averaged spectrum (a), time lags (b)
   and absolute variability amplitude for different assumptions
   regarding the radial ionization profile. For all lines, we assume
   the default parameters, and for the time lag we assume that the
   reference band was the $2-10$ keV EPIC-pn flux. For simplicity, we
   ignore the $\mu_e$ and $E_{cut}$ dependencies explored in Fig
   \ref{fig:mueandg}. The thick black lines are for a constant disk
   ionization parameter, $\log_{10}\xi=3.0$. The other lines assume a
   radial ionization profile self-consistently calculated from the
   emissivity profile and the density profile relevant to Shakura and
   Sunyaev's zone A. From bottom to top, the red, green, blue and cyan
   lines assume a `peak' ionization of $\log_{10}\xi =$3, 3.5, 3.75,
   4.0. We see that accounting for the radial ionization profile makes
   an enormous difference to the results.}
 \label{fig:xivarstressfree}
\end{figure}

Another effect to consider is the radial dependence of the incidence
angle of illuminating photons $\delta_i$ (see Fig \ref{fig:schem}),
the cosine of which is plotted for the default parameters in Fig
\ref{fig:mu} (red line). The incidence angle influences
the shape and normalisation of the restframe reflection spectrum
\citep{Garcia2010} but, in order to save computational expense, the
public \textsc{xillver} grid is tabulated only for
$\delta_i=45^\circ$. Since the leading order effect is on the
intensity of the radiation field at the disk upper boundary, $2
F_x(r)/\mu_i(r)$, the radial $\mu_i$ profile can be approximately
accounted for very cheaply by adjusting the ionization profile
(\citealt{Garcia2010,Dauser2013}). The red lines in Fig
\ref{fig:xivsr} show the logarithm of the `effective' ionization 
parameter, $\xi_{eff}(r) = \xi(r) \cos(45^\circ)/\mu_i$, that results
from this adjustment. We use this effective ionization in our model
since it captures more physics for no extra computational cost.

\begin{figure}
  \begin{centering}
	\includegraphics[angle=0,width=\columnwidth,trim= 1.7cm 1.5cm 1.8cm 3.0cm,clip=true]{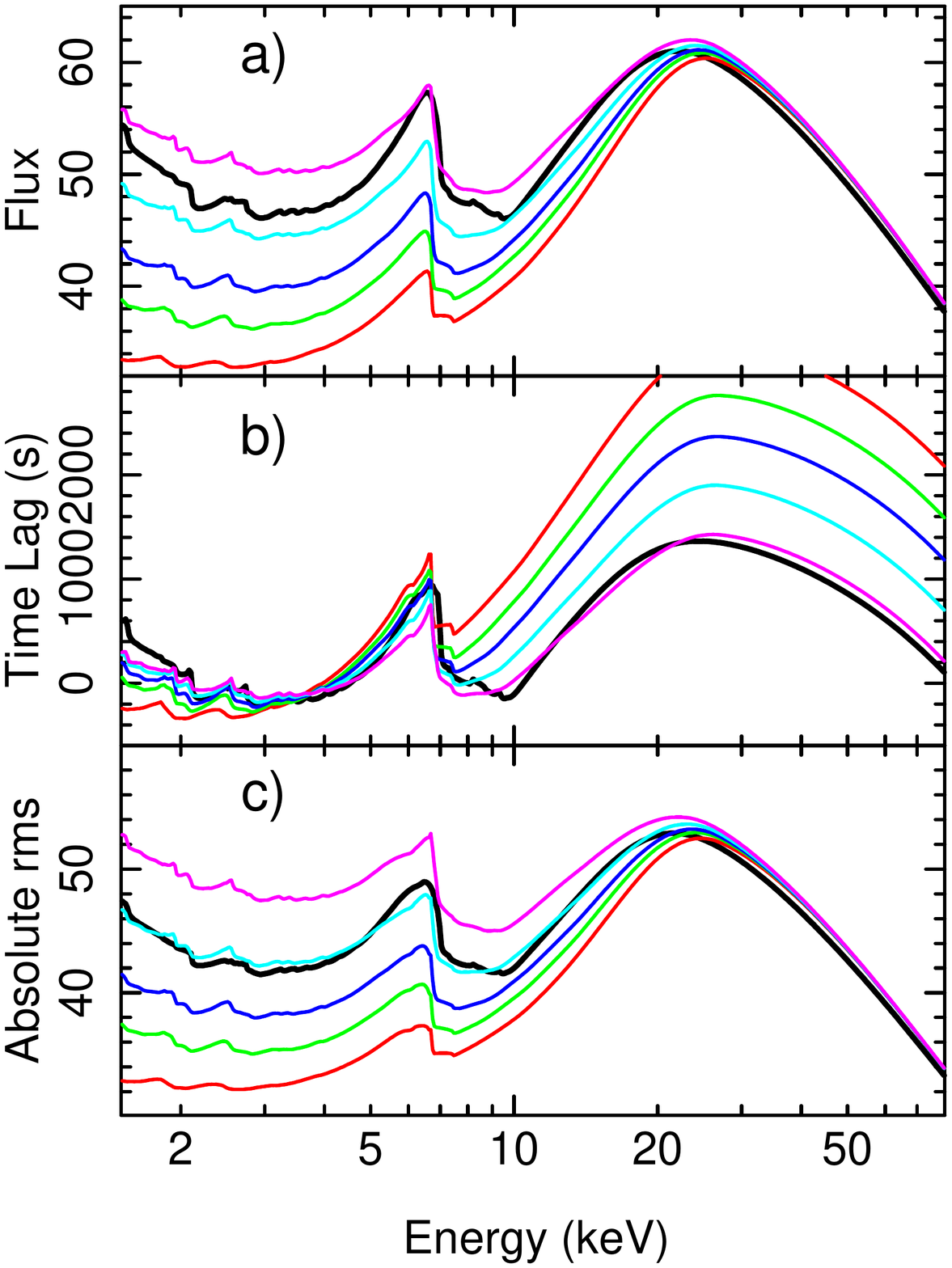}
  \end{centering}
\vspace{-8mm}
 \caption{
Same as Fig \ref{fig:xivarstressfree}, except the radial ionization
profile is calculated assuming a disk with constant density. The
magenta lines are for $\log\xi(r_{in})=4.25$. For this constant density
case, the peak ionization that most closely matches the spectrum to
the constant ionization case is slightly higher (i.e. $4.25$, magenta
line). We again see that the red wing of the iron line in the
time-averaged spectrum (a) is exaggerated compared with the constant
ionization model (thick black lines), even though the change in shape
of the line is slightly less dramatic than for Fig
\ref{fig:xivarstressfree}. The use of a self-consistent ionization
profile again has a significant affect on the time lags (b) and
absolute rms (c), with the self-consistent ionization profile again
leading to a more variable iron line (see magenta lines).}
 \label{fig:xivarncons}
\end{figure}

Figs \ref{fig:xivarstressfree} and \ref{fig:xivarncons} show the
time-averaged spectrum (a), time lag (b) and amplitude (c). The thick
black lines are computed for a single ionization parameter of
$\log_{10}\xi=3$, whereas a self-consistently calculated effective
ionization profile has been used for the coloured lines. We use the
zone A density profile for Fig \ref{fig:xivarstressfree} and constant
density for Fig \ref{fig:xivarncons}. From bottom to top, the red,
green, blue, cyan and magenta lines are for $\log_{10}\xi_{max}=$ 3,
3.5, 3.75, 4.0 and 4.25 respectively. We see that this modification to
the model makes an enormous difference to all outputs. For the zone A
profile, setting $\log_{10}\xi_{max}=3.75$ (blue lines) gives the
closest match to the constant ionization model in terms of the
relative peak fluxes of the time-averaged iron line and reflection
hump. However, the red wing of the iron line is much more prominent
for the self-consistent case. The constant density case is similar,
except for $\log_{10}\xi_{max}=4.25$. We will investigate possible
biases that this may cause when fitting constant ionization models to
observational data in section \ref{sec:ionbias}. The effect is even
greater if we also consider the timing properties. The self-consistent
models have smaller iron line lags than the single ionization model,
which may perhaps be mistaken for the source height or disk inner
radius being smaller. The absolute rms spectrum shows that the iron
line is much more variable for the self-consistent case. As we will
see in section \ref{sec:ionbias}, it is possible to choose a value of
$h$ that allows the single ionization model to mimick the
time-averaged spectrum of the self-consistent case, but the different
effect that the ionization gradient has on the timing properties means
that considering also time lags and variability amplitude breaks the
degeneracy. Both figures here are plotted using 100 zones in
ionization parameter. We find however that reasonable convergence can
be achieved for 10 zones, and so we set this as the default value for
the \texttt{ION\_ZONES} environment variable (see Table
\ref{tab:env}).

\subsubsection{\textsc{reltransCp} vs \textsc{reltrans}}
\label{sec:Cp}

\begin{figure}
  \begin{centering}
	\includegraphics[angle=0,width=\columnwidth,trim= 1.7cm 1.5cm 1.8cm 3.0cm,clip=true]{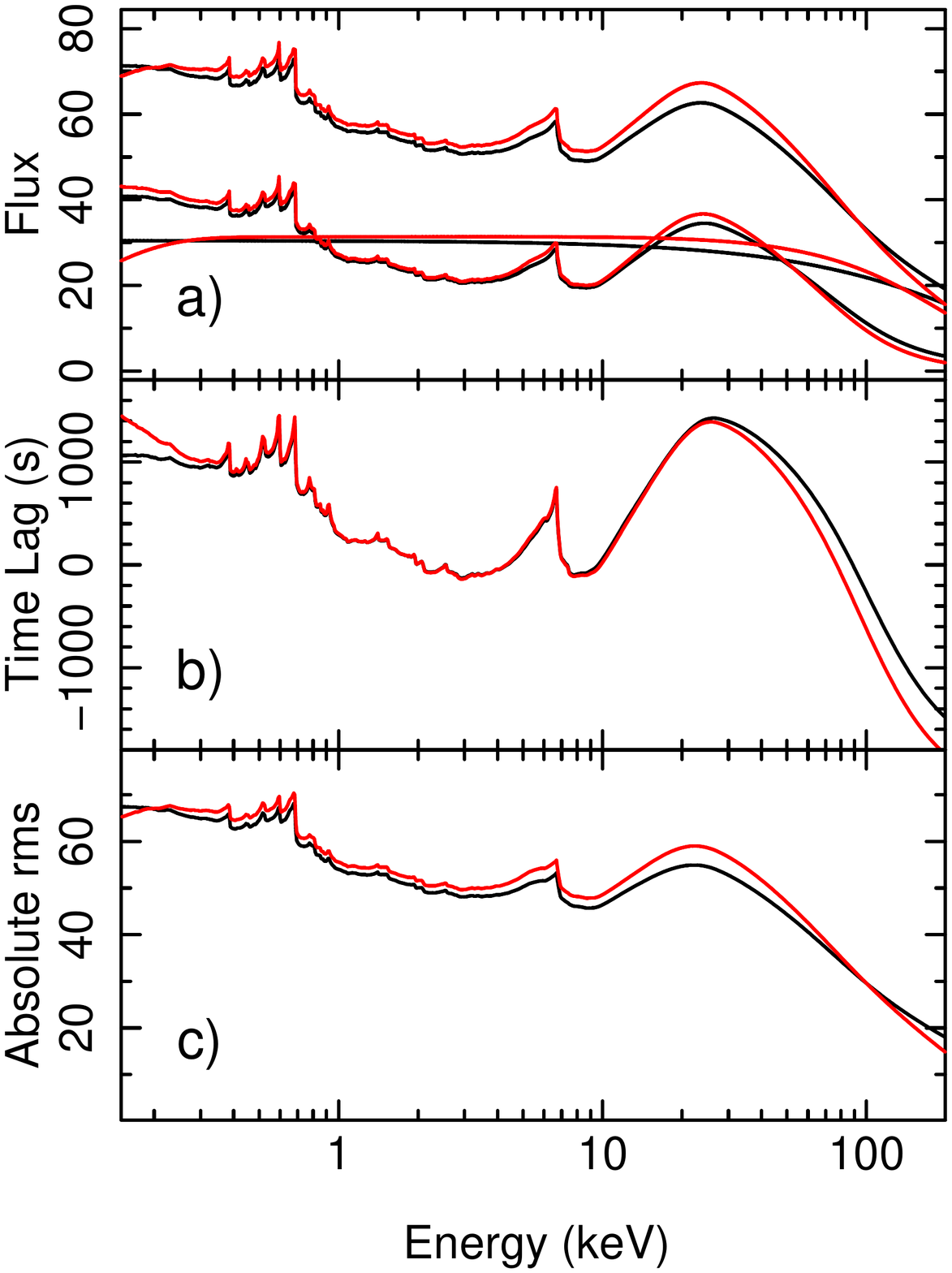}
  \end{centering}
\vspace{-8mm}
 \caption{Time-averaged spectrum (a), time lags (b) and absolute
   variability amplitude calculated from the default parameters. The
   black lines represent the model \textsc{reltrans} (which assumes an
   exponentially cut-off power-law for the direct spectral component) and the red
   lines represent \textsc{reltransCp} (which uses \textsc{nthcomp}
   for the direct spectrum). We see a significant difference in the
   time-averaged direct component. The time lags are largely unaffected for
   energies below $\sim 20$ keV, whereas the more physical model
   \textsc{reltransCp} predicts a larger variability amplitude. Here
   we use default values for the environment variables, and the phase
   zero point is self-consistently calculated for a $2-10$ keV \xmm~pn
   timing mode reference band.}
 \label{fig:Cp}
\end{figure}

Fig \ref{fig:Cp} demonstrates typical differences between
\textsc{reltrans} (black lines) and \textsc{reltransCp} (red
lines). Whereas the former uses an exponentially cut-off power-law
for the illuminating spectrum, the latter uses the model \textsc{nthcomp}
(\citealt{Zdziarski1996,Zycki1999}), which gives a much better
approximation of Compton up-scattering of seed photons by a thermal
population of hot electrons. We see that \textsc{nthcomp} has a
low energy cut-off, which is determined by the seed photon temperature
$kT_{bb}$. In the \textsc{xillverCp} tables, this is hardwired to
$0.05$ keV (assuming a multi-temperature blackbody spectrum of seed
photons). The shape of the high energy cut-off is also very different
for \textsc{nthcomp}. The difference between the two models is small
for the lags though. Since \textsc{reltransCp} employs a more physical
emission model for little extra computational expense, we use it for
the remainder of the plots in this paper.

\subsection{Frequency dependence}
\label{sec:freq}

\begin{figure}
  \begin{centering}
	\includegraphics[angle=0,width=\columnwidth,trim=1.0cm 1.5cm
        2.5cm 16.2cm,clip=true]{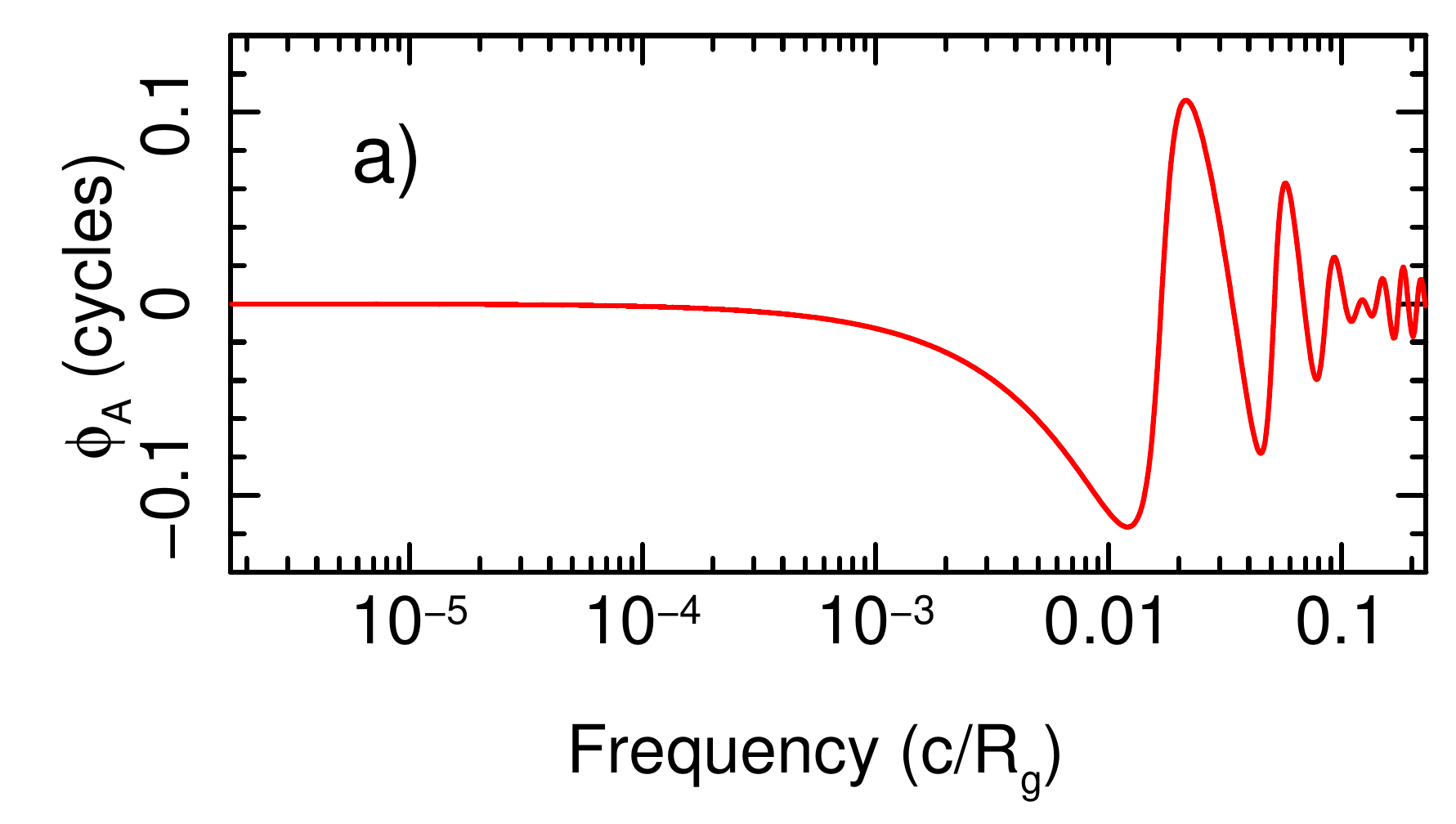} \\
	\includegraphics[angle=0,width=\columnwidth,trim= 1.7cm 1.5cm 2.5cm 9.5cm,clip=true]{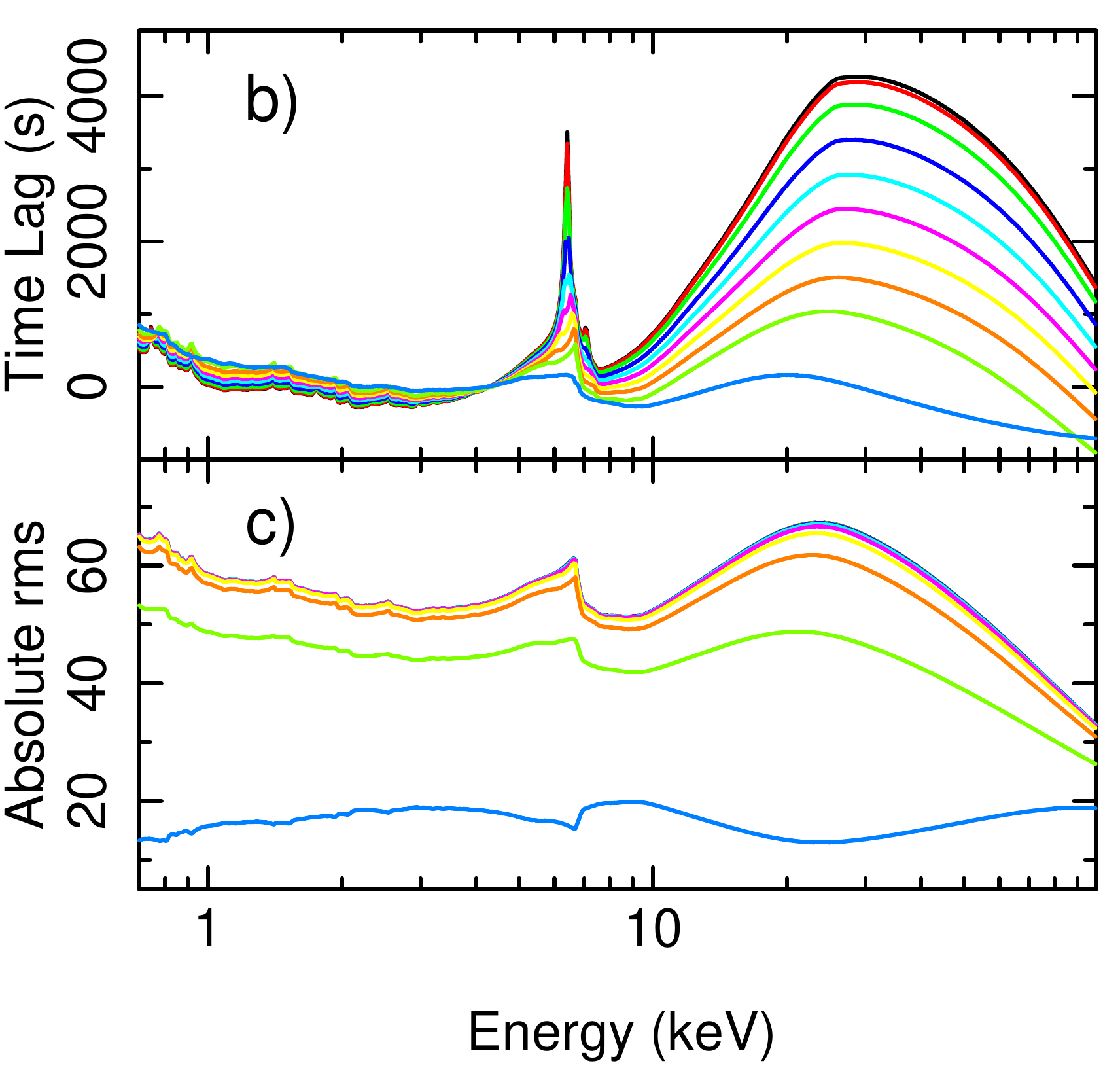}
  \end{centering}
\vspace{-5mm}
 \caption{
\textit{a:} $\phi_A(\nu)$ calculated using the default model
parameters of \textsc{reltransCp} (default environment variables),
assuming that the reference band is the $2-10$ keV EPIC-pn
flux. \textit{b:} Time lag versus energy
for the same parameters using the same calculation for
$\phi_A(\nu)$. The black, red, green, blue, cyan, magenta, yellow,
orange, light green and light blue lines (top to bottom) are
calculated for different frequency ranges, which increase
logarithmically from a minimum of $10^{-8}$ Hz to a maximum of
$10^{-4}$ Hz (this range corresponds to $0.046-460$ Hz for a $10
M_\odot$ black hole, or $\sim 2.26 \times
10^{-6}-10^{-2}~c/R_g$). \textit{c:} The same calculation but for
variability amplitude, setting $\alpha(\nu)=1$.}
 \label{fig:freq}
\end{figure}

Fig \ref{fig:freq} demonstrates the frequency dependence of
\textsc{reltransCp} for the default parameters. Panel a shows the phase
normalisation $\phi_A(\nu)$ calculated for a $2-10$ keV reference
band measured by the EPIC-pn instrument onboard the \textit{X-ray
  Multi-mirror Mission} (\xmm; \citealt{Jansen2001}) in timing mode
(calculated from equation \ref{eqn:phiA}). As noted in
\cite{Mastroserio2018}, we can only be confident that this function is
a correct representation of the underlying spectral model if all the
channels used for the reference band are considered to be well
calibrated. The range $2-10$ keV demonstrated in the plot is well
calibrated for \xmm. If for any reason we wish to define our reference
band from poorly calibrated channels, for instance if we wish to
maximize signal to noise by collecting more photons, then a systematic
error will be introduced into the calculation of $\phi_A(\nu)$ because
the instrument response matrix used for the calculation does not
adequately describe the true response of the telescope for all energy
channels. In such a case, it may be best to leave $\phi_A(\nu)$ as a
free parameter for each frequency range considered, although this will
inevitably lead to larger statistical errors.

Panels b and c show respectively time lags and absolute variability
amplitude as a function of energy for 10 different frequency
ranges. The overall lag reduces and the iron
line feature gets broader with increasing frequency because the higher
frequencies select reflection from smaller regions of the
disk. Similarly, the line feature in the rms spectrum becomes weaker
for higher frequencies as the fastest variability is washed out by
path length differences introduced by reflection from different parts
of the disk. At the highest frequency range plotted here, we see the
effects of phase-wrapping, evidenced by the iron line and reflection
hump becoming dips as opposed to excesses in the rms
spectrum.

Our model calculates the energy dependent cross-spectrum for a given
frequency range, rather than the cross-spectrum as a function of
frequency for a given energy range. This feature is hardwired because
we calculate the energy dependent transfer function in Fourier space
(see equation \ref{eqn:transo}) for a range of frequencies between
$\nu_{min}$ and $\nu_{max}$. This is much more computationally
efficient than first calculating a 2D impulse-response function and
Fourier transforming the time axis. All frequencies can be taken into
account by fitting for many frequency ranges, as in
\cite{Mastroserio2018}. The public model does not currently include
non-linear effects, but will soon be updated (description in
Mastroserio et al submitted). Intrinsic hard lags can alternatively be
produced by summing two model components. This is only possible when
considering real and imaginary parts of the cross-spectrum, and the
phase normalisations of the two component must be free parameters
(i.e. not self-consistently calculated) and independent of one
another. Using, for example, two \textsc{reltrans} components with
different source heights and different spectral indices is the same as
assuming the `two blobs' geometry of \cite{Chainakun2017}. The Fourier
frequency dependent propagation time between the blobs is simply
$|\phi_{A1}(\nu)-\phi_{A2}(\nu)|/(2\pi \nu)$.

\section{Modelling biases}
\label{sec:bias}

In this section, we explore two sources of bias in previous treatments
of reflection and reverberation in the literature. The first is
ignoring the instrument response matrix when analysing time lags. In
section (\ref{sec:lagbias}), we show that the value of the lag can be
heavily biased in an energy range for which the instrument response is
not diagonal and in which line-of-sight absorption is prominent. This
is because such an energy range is \textit{dominated} by photons from
other energy bands that have been `mis-classified'. The other source
of bias is assuming a single disk ionisation parameter instead of
accounting for a self-consistently calculated radial ionisation
profile (section \ref{sec:ionbias}).

\subsection{Bias caused by ignoring the telescope response}
\label{sec:lagbias}

Our model provides two ways to properly account for the instrument
response and line-of-sight absorption. The recommended option for the
purposes of fitting to data is to consider the real and imaginary
parts of the cross-spectrum (\texttt{ReIm}=1 and 2). In this case, the
fits files containing the data can be read into \textsc{xspec} in the
normal way, with response files specified in the header, and the usual
\textsc{xspec} operation of folding around the instrument response is
appropriate. The user can then plot the best fitting cross-spectral
model in terms of variability amplitude and time lags after the fit is
complete (\texttt{ReIm}=3 and 4). However, the user may wish to
instead fit for time lags and/or variability amplitude. In this case, the
model cross-spectrum is folded around the instrument response within
the code and the variability amplitude and time lags are calculated
from this folded cross-spectrum (\texttt{ReIm}=5 and
6). Observationally constrained time lags and rms can be loaded into
\xspec~with a diagonal dummy response matrix (using
e.g. \texttt{flx2xsp}). The response files can be set through
environment variables (\texttt{RMF\_SET} and \texttt{ARF\_SET}). If
the environment variables are not set but the code is in a mode
requiring a response, the user will instead be prompted at the
terminal for input.

\begin{figure}
  \begin{centering}
	\includegraphics[angle=0,width=\columnwidth,trim=1.5cm 1.5cm
        2.5cm 11.0cm,clip=true]{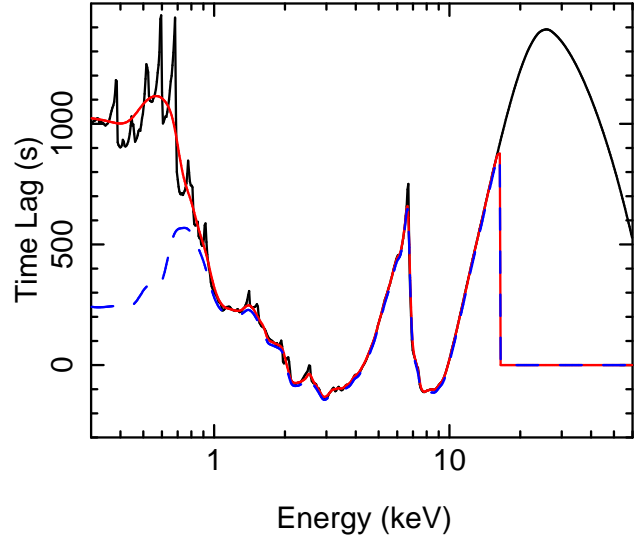}
  \end{centering}
\vspace{-5mm}
 \caption{Time lags calculated for the default parameters. The black
   line is for the model only, and so does not account for a telescope
   response matrix. The red and blue lines assume the pn response and
   absorption column densities of $N_h=0$ and $10^{22}{\rm cm}^{-2}$
   respectively. We calculate $\phi_A$ for all three assuming the
   reference band is the $2-10$ keV band of the pn.}
 \label{fig:abs}
\end{figure}

\begin{figure}
  \begin{centering}
	\includegraphics[angle=0,width=\columnwidth,trim=1.5cm 1.5cm
        2.5cm 11.0cm,clip=true]{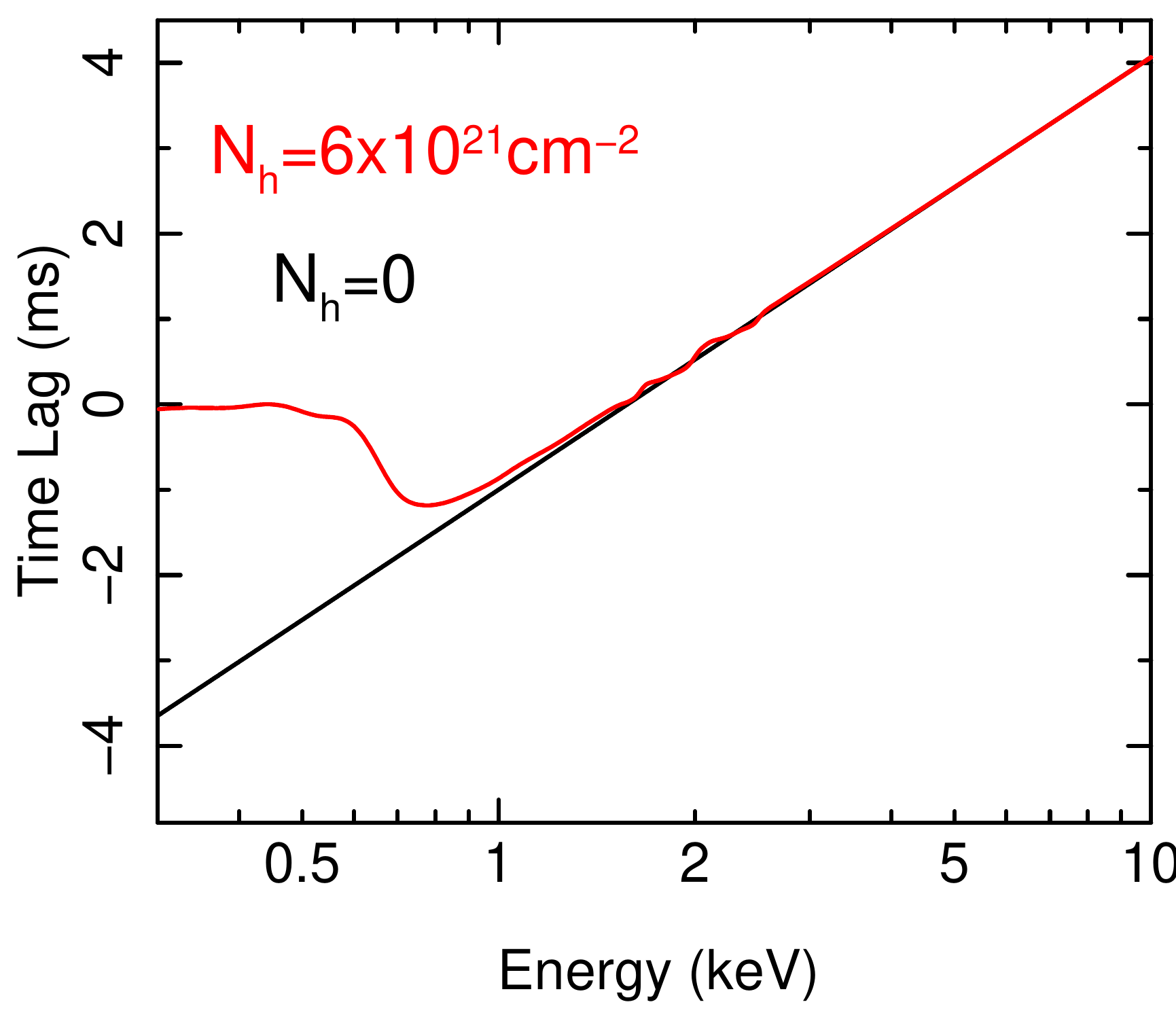}
  \end{centering}
\vspace{-5mm}
\caption{Log-linear time lag as a function of energy (black), with the
  parameters chosen to roughly match the $1-30$ Hz lag spectrum of GX
  339-4 in observation `O1' from \citet{DeMarco2017} and
  \citet{Mahmoud2019}. The red line is the time lag that would be
  observed by \textit{XMM-Newton} assuming that the intrinsic
  lag-energy spectrum is given by the black (log-linear) line. To
  calculate this, we take the argument of the absorbed and folded
  model cross-spectrum. We see that an intrinsically log-linear lag
  spectrum appears to turn up at $E \lesssim 1$ keV in
  \textit{XMM-Newton} observations.}
 \label{fig:inst}
\end{figure}

Fig \ref{fig:abs} illustrates the importance of correctly accounting
for the instrument response. The black line represents time lags for
the model only, ignoring the instrument response
(i.e. \texttt{ReIm}=4). In this case, the absorption model is not
relevant because it cancels out when the imaginary part of the
cross-spectrum is divided by the real part. The red line represents
the same model, but now the EPIC pn response has been used and we set
the hydrogen column density to $N_h=0$. The blue dashed line is the
same again except that now we set $N_h = 10^{22} {\rm cm}^{-2}$. We
see very little differences in the $2-10$ keV region, whereas above
$\sim 15$ keV the lags are completely undefined due to lack of
effective area. The differences between the lines below $\sim 1$ keV
occur because the response matrix is not diagonal in this energy
range. For $N_h=0$, the resulting ambiguity between soft photons and
harder photons `mis-classified' as soft simply smears out sharp
features. When absorption is taken into account, the soft band instead
becomes \textit{dominated} by the mis-classified photons. This
essentially introduces dilution: the lag between $0.5$ and $1$ keV is
very small because most of the photons recored in the $\sim 0.5$ keV
channel are actually $\sim 1$ keV photons. 

Whereas Fig \ref{fig:abs} is relevant for high frequencies at which
the reverberation lags dominate over the intrinsic hard lags, the same
effect is also potentially important for lower frequency ranges in
which the intrinsic lags are still significant. In particular,
signatures of thermal reverberation have been detected for a number of
black hole X-ray binaries including GX 339-4
\citep{Uttley2011,DeMarco2017}. In the $1-30$ Hz frequency range,
log-linear intrinsic lags are seen for $E \gtrsim 1$ keV and a turn up
is seen for $E \lesssim 1$ keV, which is attributed to thermal
reverberation (see top right of Fig 7 in \citealt{DeMarco2017}). We
investigate how this thermal reverberation signal may have been
affected by the instrument response by first assuming that the
intrinsic lag spectrum in the $1-30$ Hz frequency range is simply
log-linear for the full energy range (Fig \ref{fig:inst}, black
line). From this, we calculate the energy dependent
cross-spectrum. This additionally requires a model for the
time-averaged spectrum and a model for the energy dependent fractional
variability amplitude. We use \textsc{tbabs}*\textsc{nthComp}
($N_h=6\times 10^{-21} {\rm cm}^{-2}$, $\Gamma=1.9$, $kT_{bb}=0.18$
keV) for the time-averaged spectrum and assume that the fractional rms
increases linearly with energy ($rms \propto 0.024 E + 0.043$,
although our results only depend very weakly on this function). We
then fold our model cross-spectrum around the \textit{XMM-Newton} pn
timing mode response matrix and calculate the `folded' lag spectrum
from the argument of this folded cross-spectrum (red line). We see a
clear turn up in the `folded' lag spectrum below $\sim 1$ keV that
results from these energy channels being dominated by `mis-classified'
photons.

On first inspection, this looks worryingly like a spurious signature
of thermal reverberation. However, the observation of GX 339-4 that
our model is based upon has a number of characteristics that
convincingly point to the presence of thermal reverberation. In
particular, \cite{DeMarco2017} present the $5-30$ Hz lag energy
spectrum in their Fig 7. That the $\sim 0.5$ keV lag increases as
progressively higher frequency ranges are chosen is very suggestive of
thermal reverberation. Moreover, the $E \sim 0.5$ keV lag is
\textit{larger} than the lag in the $\sim 1-6$ keV energy range, and
so simply cannot be caused by the instrumental effect that we have
explored here - which can only dilute the soft lags by averaging with
the higher energy lags. We therefore conclude that the thermal
reverberation interpretation of the data is sound. However, the
\textit{value} of the lag is very likely biased by the instrument
response. In particular, \cite{Mahmoud2019} fit a transfer function
model to the data that only accounts for the effective area curve of
the instrument but not the redistribution matrix. This implies that
the true reverberation lags are shorter than originally thought, and
the measured disk inner radius of $\sim 20~R_g$ may reduce once the
correction is made.

We conclude that it is important to properly account for the telescope
response matrix when modelling the $\lesssim 1$ keV region of
\xmm~data, either by fitting for real and imaginary parts, or using
the folding option. A similar effect is present in the
\nicer~response, but not that of the $\textit{Nuclear Spectroscopic
  Telescope ARray}$ \citep[\nustar;][]{Harrison2013}.

\subsection{Bias caused by using a single ionization}
\label{sec:ionbias}

It is clear from the discussion in Section \ref{sec:details}, and in
particular Figs \ref{fig:xivarstressfree} and \ref{fig:xivarncons},
that including a self-consistent radial ionization profile can give
very different model outputs to simply assuming a constant ionization
parameter. In this section, we create a fake \nustar~time-averaged
spectrum and fit back with a single ionization parameter model in
order to investigate biases that may have been introduced into the
many spectral fitting studies that have used a single ionization
model.

\subsubsection{Fake data}

We simulate a $30$ ks \nustar~observation of a bright X-ray binary
by inputing the model parameters listed in Table \ref{tab:fitback}
into \textsc{reltransCp}, with the normalisation set to roughly match
the observed flux of GX 339-4 when $\Gamma \approx 2$ (model $4-10$
keV flux $=5\times 10^{-9} $ erg~cm$^{-2}$~s$^{-1}$). We use
\textsc{fakeit} to generate a fake $30$ ks FMPA exposure, taking
background into account. We ignore deadtime effects, but do not
generate an FPMB exposure. Statistically, this is the same as taking
both focal plane modules into account and assuming a deadtime
correction factor of $1/2$. Our fake observation therefore corresponds
to a typical high quality observation used for the purposes of
spectral fitting (e.g. \citealt{Parker2015,Xu2018,Tomsick2018}). We
only simulate the time-averaged spectrum, since there have thus far
only been a few studies fitting reverberation models to timing data.

\begin{figure*}
  \begin{centering}
	\includegraphics[angle=0,width=\columnwidth,trim=1.0cm 1.5cm
        3.0cm 2.1cm,clip=true]{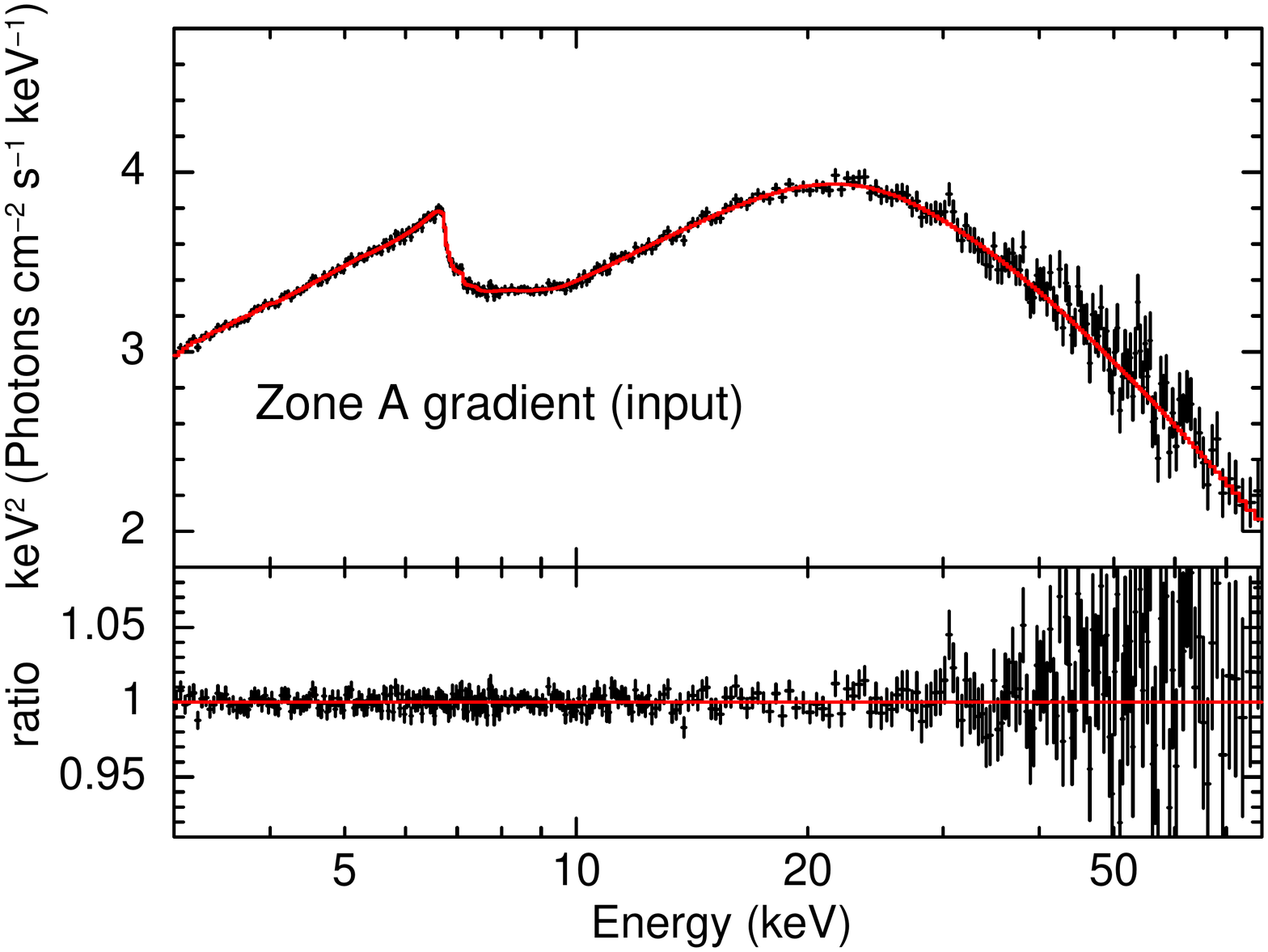} ~~~~
	\includegraphics[angle=0,width=\columnwidth,trim=1.0cm 1.5cm
        3.0cm 2.1cm,clip=true]{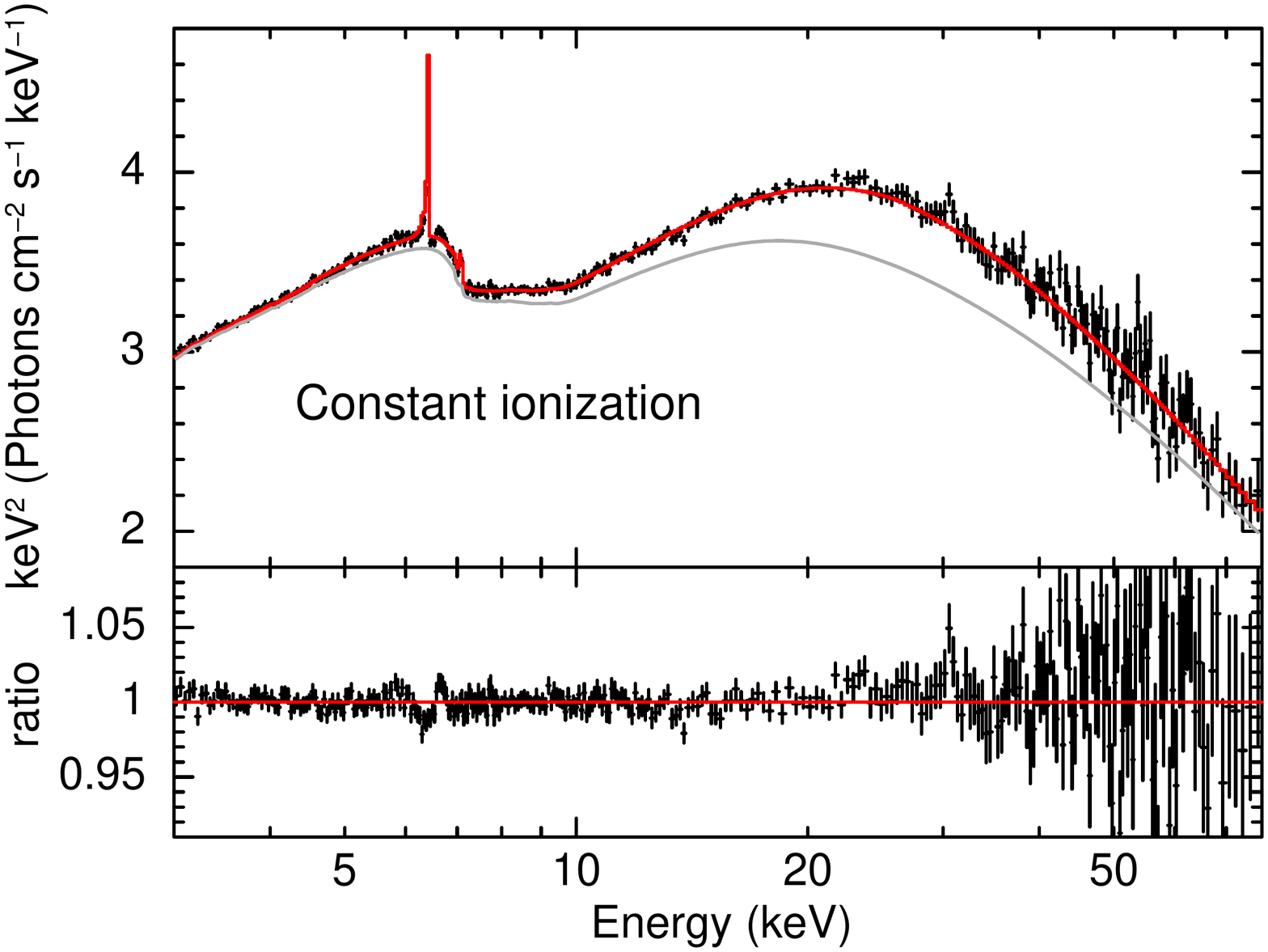}
  \end{centering}
\vspace{0mm}
 \caption{Unfolded fake data and model (top) and fake data to model
   ratio (bottom). The data were generated from a model with a radial
   ionization profile calculated assuming a zone A Shakura-Sunyaev
   density profile. Left and right hand plots respectively show the
   results of fitting the input model and an alternative model with a
   single ionization parameter (red) plus an extra
     \textsc{xillver} component (grey). Results are further
   detailed in Table \ref{tab:fitback}. Spectra have been re-binned
   for plotting purposes (binned to a target signal to noise ratio of
   150 but not co-adding more than 10 channels).}
 \label{fig:fitback}
\end{figure*}

\begin{table}
\begin{center}
\renewcommand{\arraystretch}{1.5}
\begin{tabular}{|l|l|l|l|} 
\hline
Parameter & Input & Control Fit & Single Ion Fit \\
\hline
\hline
$N_H$ ($10^{22}{\rm cm}^{-2}$) & 1 & 1 & 1 \\
\hline
$h$ ($R_g$) & 6 & $5.48_{-0.546}^{+0.616}$ & $2.45_{p}^{+0.26}$ \\
\hline
$a$  & 0.9 & $0.99_{-0.099}^{p}$ & $0.778_{-0.0581}^{+0.0461}$ \\
\hline
$i$ (degrees)  & 30 & $30.7_{-1.72}^{+1.06}$ & $29.1_{-1.12}^{+0.89}$ \\
\hline
$\Gamma$  & 2 & $2.0_{-0.01}^{+0.01}$ & $2.03_{-0.009}^{+0.015}$ \\
\hline
$\log_{10}\xi^*$  & 3.75 & $4.13_{-0.587}^{+0.289}$ & $3.05_{-0.034}^{+0.034}$ \\
\hline
$A_{Fe}$ & 1 & $0.967_{-0.1259}^{+0.1926} $ & $0.999_{-0.0599}^{+0.1198}$ \\
\hline
$(kT_e)_{obs}$ (keV)  & 50 & $46.1_{-5.37}^{+15.0}$ & $50.8_{-3.23}^{+11.81}$ \\
 \hline
$1/\mathcal{B}$  & 0.5 & $0.571_{-0.0743}^{+0.1694}$ & $0.419_{-0.0188}^{+0.0511}$ \\
 \hline
norm ($10^{-2}$) & 10 & $7.97_{-0.621}^{+0.959}$ & $56.8_{-12.51}^{+4.86}$ \\
 \hline
\textsc{xillverCp} norm ($10^{-3}$) & 0 & 0& $6.32_{-0.078}^{+0.08}$ \\
\hline
\hline
$\chi^2/$d.o.f.   &   & $1540.68/1561$ & $1505.89/1560^{**}$ \\
\hline
\end{tabular}
\renewcommand{\arraystretch}{1}
\end{center}
\caption{Input parameters and values measured by fitting back with the
  correct (control) model and a single ionization model. Errors are
  $90\%$ confidence limits, and $p$ denotes that the parameter is
  pegged at a hard limit. $^*$The ionization parameter has different meanings in
  the two models (single value vs maximum value). $^{**}$ A $0.5\%$
  systematic error was applied for the single ionization model fit, as
  is commonly practiced in spectral fitting to account for calibration
  uncertainty and model systematics. Without the systematic error, the
best fitting single ionization model has a reduced $\chi^2$ of
1626.38/1560, corresponding to a goodness of fit that is common for
fits to real \textit{NuSTAR} spectra \citep[e.g.][]{Miller2013}.}
\label{tab:fitback}
\end{table}

Our input model assumes the Zone A Shakura-Sunyaev density
profile. This is a reasonable assumption for the brightest hard / hard
intermediate states of X-ray binaries. The bolometric luminosity of GX
339-4 is $L_x \sim 10^{38.5}$ erg s$^{-1}$ when $\Gamma=2$ (see Fig 5
of \citealt{Plant2014}).
For $M \lesssim 10~M_\odot$ 
(\citealt{Heida2017}) and a viscosity parameter $\alpha \gtrsim 0.01$, the zone
A to B transition is therefore at $r_{ab} \gtrsim 200~R_g$ (equation
2.17 in \citealt{Shakura1973}), indicating that the 
region of the disk that dominates the emissivity is radiation pressure
and electron scattering dominated. Following the discussion in section
\ref{sec:details} concluding that the disk coordinate dependence of
$\mu_e$ and disk rest frame observed electron temperature are not
important for low source inclinations, we use 100 ionization zones for
our input model, and only one for the emission angle and electron
temperature.

\subsubsection{Fit results}

Fig \ref{fig:fitback} and Table \ref{tab:fitback} summarise the
results of fitting the fake data with the input model (left) and a
single ionization model (right). We fix the hydrogen column density in
both of the fits, assuming this to  be constrained in some other
way. We see that the single ionization model under-predicts the source
height with high statistical significance. This is consistent with
\cite{Svoboda2012} and \cite{Kammoun2019}, who found that using a
single ionization zone can produce artificially steep power-law
emissivity profiles (and lower source height corresponds to steeper
emissivity: Fig \ref{fig:emissivity}). Previous spectral fitting studies using
lamppost models assuming a single ionization parameter have therefore
likely under-predicted the source height. The assumption of a single
ionization parameter seems to have introduced a small bias in the spin
measurement, although we find that the spin is under-predicted here,
whereas many observational studies, particularly for AGN, yield
near-maximal spin values. We also note that the disk inner radius is
fixed to the ISCO in our fake data, but is often observed to reduce as the
spectrum softens in the real data
(e.g. \citealt{Plant2015,Garcia2015}). The large red wing of the iron
line introduced by the ionization gradient appears to have been
compensated by a larger value of $\Gamma$ instead of a smaller value
of $r_{in}$.

Interestingly, the single ionization fit includes a highly
statistically significant ($5.5~\sigma$ from an F-test) low-ionization
\textsc{xillverCp} component. Such a component is often required in
fits to real spectra in order to account for enhanced distant
reflection (e.g. from a flared outer disk, or from the companion
star). Our experiment here implies that the often uncomfortably high
flux required for the distant reflector may, in part, be due to a
modelling systematic introduced by assuming a single ionization
parameter. The correct iron abundance is recovered for both fits, but
we note that the best fitting single ionization model with no distant
reflection component includes a super-solar iron abundance
($A_{Fe}=2.71_{-0.229}^{+0.313}$; errors are $90\%$
confidence limits). This is interesting because super-solar iron
abundances are now consistently measured in X-ray binaries, but it is
not well understood why this should be the case
(\citealt{Garcia2018}). It is suspected that the iron abundance
parameter is compensating for some missing physics in the models, such
as higher electron density in the disk ($n_{\rm e} \sim 10^{20-22}
{\rm cm}^{-3}$; \citealt{Tomsick2018}). The
assumption of a single ionization parameter may also contribute to the
high measured iron abundance in some cases.

\begin{figure*}
        \includegraphics[width=0.48\textwidth]{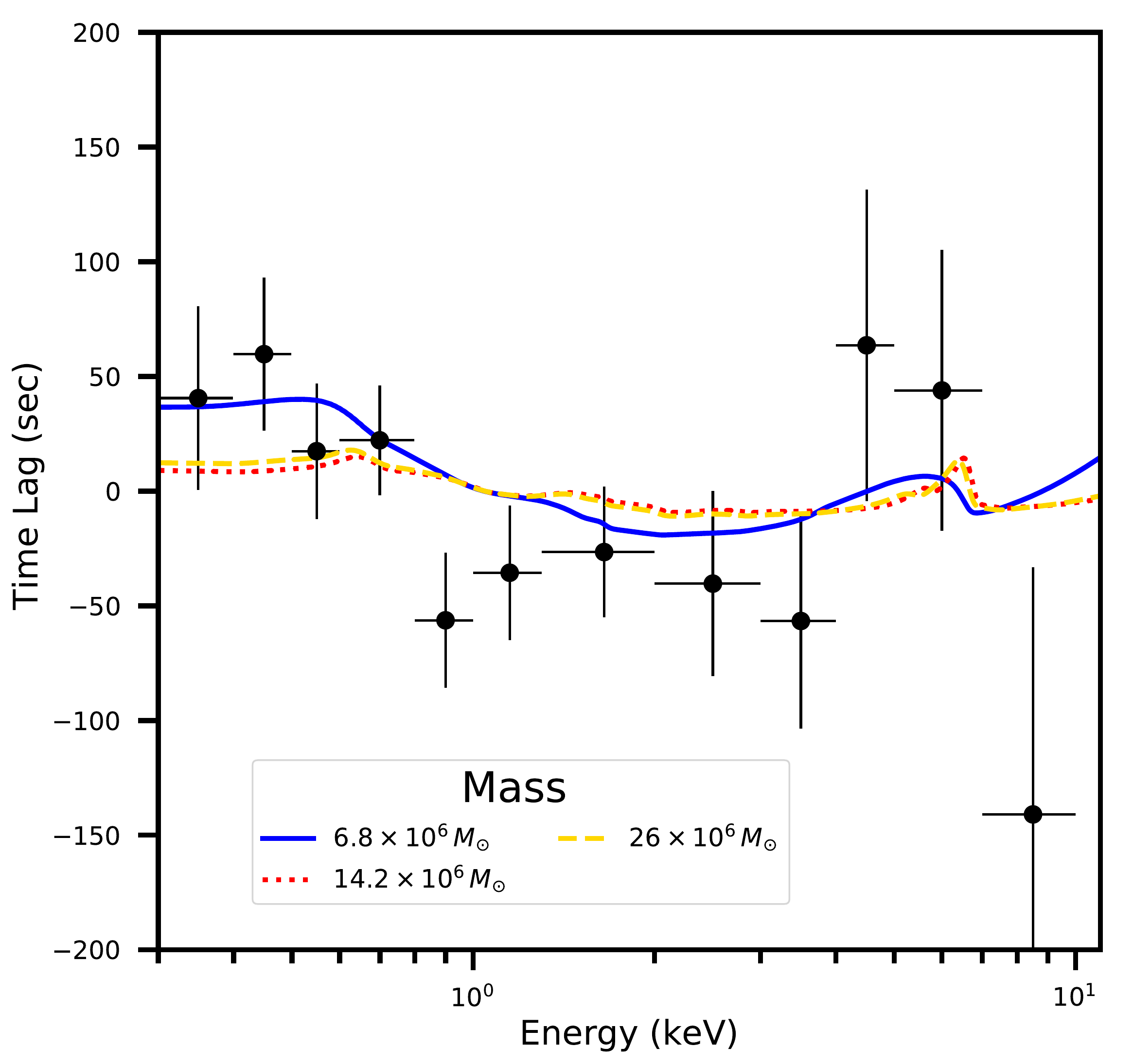}
  ~~~~
        \includegraphics[width=0.48\textwidth]{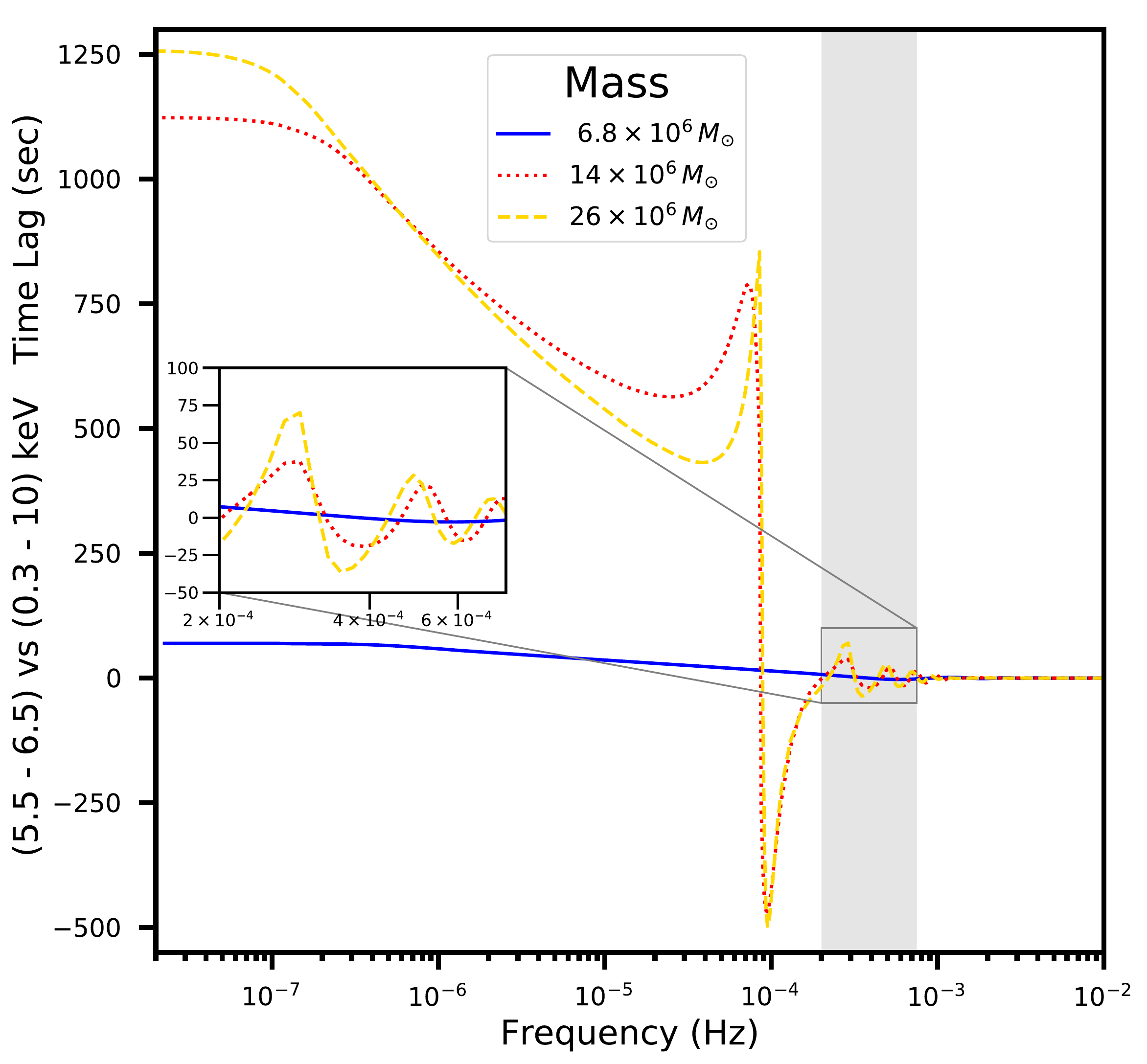}
    \centering
    \caption{\textit{Left:} Time lag as a function of energy in the
      frequency range $[2 {\rm -} 7.5] \times 10^{-4}$ Hz for
      \textit{XMM-Newton} data from Mrk 335 (black points), alongside
      three \textsc{reltrans} model fits (reference band: $0.3-10$
      keV). For the blue solid line, the black hole mass is a free
      parameter, and for the red dotted and yellow dashed lines we fix
      it to two different optical reverberation values from the
      literature. \textit{Right:} Time lag, averaged over the iron
      line region ($5.5-6.5$ keV), plotted against frequency for the same three
      models. We see that, in the frequency range used for the
      lag-energy spectrum (grey band), the models using the optical
      reverberation masses (red dotted and yellow dashed lines) are in
      the phase-wrapping regime whereas our best fitting model (blue
      solid line) is not.}
    \label{fig:fit_mrk335}
\end{figure*}

\section{Example fits for Mrk 335}
\label{sec:Mrk335}

As a proof of principle, we apply the \textsc{reltrans} model to
an archival \textit{XMM-Newton} observation of the AGN Mrk 335 for
which an iron K feature has previously been identified in the
lag-energy spectrum \citep{Kara2013}. We choose this observation
  because it provides a good example of an iron K lag feature without
  the need for complications such as stacking multiple observations or
  dealing with photon pile-up (both of which are required for the Ark
  564 lag-energy spectrum also featured in \citealt{Kara2013}). We
  find that, even though this frequency range displays clear signs of
  reverberation, a statistically acceptable fit to the real and
  imaginary parts of the cross-spectrum could only be achieved by
  including non-linear variability of the direct spectrum, which is
  beyond the scope of this paper but is introduced for the
  \textsc{reltrans} model in a companion paper (Mastroserio et al
  2019). We therefore instead fit only the time lags in a single
frequency range here, leaving a multi-frequency fit of real and
imaginary parts of the cross-spectrum to a future paper.

\subsection{Data}

We consider the 133 ks \textit{XMM-Newton} observation taken in 2006
(obs ID  0306870101) that was analysed by \citet{Kara2013}. Following
\cite{Kara2013}, we consider only pn data, and reduce it using the
\textit{XMM-Newton} Science Analysis System (SAS v.11.0.0),  applying
the filters \texttt{PATTERN}~$\le 4$ and \texttt{FLAG}~$==$~0. We
exclude background flares at the beginning and end of the observation
(considering only times $252709714$ to $252829414$ seconds) and
extract light curves with $10$ second binning from $12$ different
energy bands, spaced roughly equally in the range $0.3-10$ keV, from a
circular region with 35 arcsec radius centred on the maximum of the
source emission. We apply the SAS task \texttt{epiclccorr} for
background subtractions and various corrections.

Again following \cite{Kara2013}, we calculate the cross-spectrum
between each of the $12$ energy bands and a reference band that is the
sum of all energy bands except for the current subject band (thereby
ensuring statistical independence between the subject and reference
bands; see e.g. \citealt{Uttley2014}). We average these $12$
cross-spectra over the frequency range $[2 {\rm -} 7.5] \times
10^{-4}$ Hz, since this is the range for which `soft lags' are observed
\citep{DeMarco2013}: i.e. fluctuations in the $0.3-0.8$ keV band lag
behind those in the $1-4$ keV band (with the former assumed to be more
reflection-dominated than the latter). We calculate energy dependent
time lags by taking the argument of each frequency-averaged
cross-spectrum and dividing by $2\pi \nu$, where $\nu=4.75\times
10^{-4}$ Hz is the centre of the frequency range. We calculate error
bars using the analytic formula from \citet[][see also
\citealt{Nowak1999}]{Bendat2010}. Since the frequency resolution
of the cross-spectra is $d\nu=8.35\times 10^{-6}$ Hz, the $[2 {\rm -}
7.5] \times 10^{-4}$ Hz frequency range contains 65 frequency bins,
meaning that the lag spectrum is Gaussian distributed and we can
therefore fit models using the $\chi^2$ statistic.

\begin{figure}
        \includegraphics[width=\columnwidth]{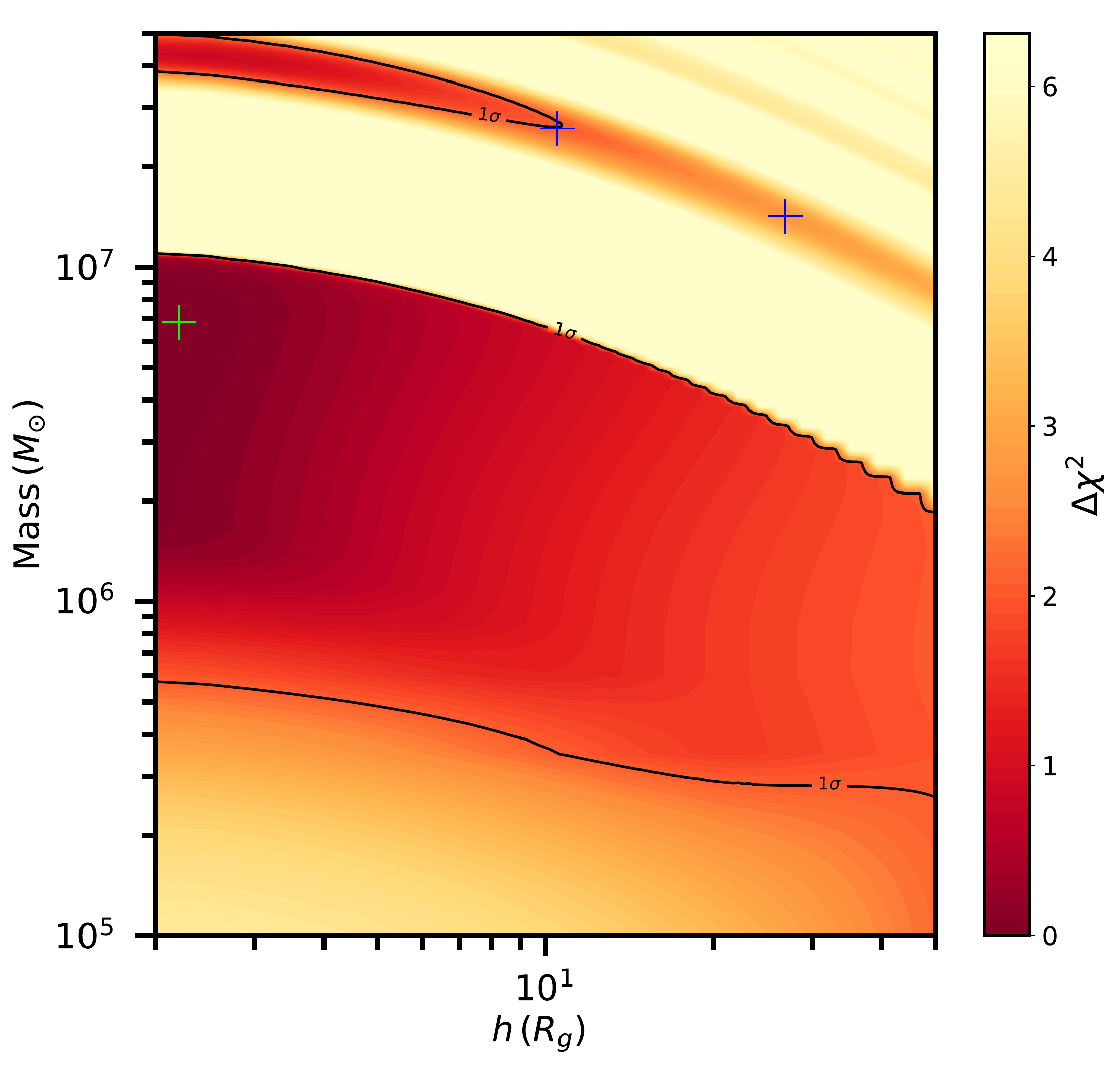}
    \centering
    \caption{2D $\chi^2$ contour plot of black hole mass and source height
      resulting from fitting \textsc{reltrans} to the lag versus energy
      spectrum of Mrk 335 in the frequency range $[2-7.5]\times
      10^{-4}$ Hz. The green cross marks the best fit and the blue
      crosses mark the best fit for for the two optical reverberation
      masses.}
    \label{fig:cont_mrk335}
\end{figure}

\subsection{Fits to the lag-energy spectrum}

Fig. \ref{fig:fit_mrk335} (left) shows the observed lag-energy
spectrum of Mrk 335 (black points), which displays the iron K feature
at $\sim 6.4$ keV reported by \cite{Kara2013}\footnote{Although note
  that we use a slightly different frequency range, and so our results
  are consistent but not identical.}, alongside three \textsc{reltrans}
model fits. A full treatment would employ a simultaneous fit to
  the time-averaged spectrum, which we will present in a future
  paper. For the purposes of this demonstration of the use of the
  model we instead fit only the lag spectrum, and avoid over-fitting
by fixing most parameters to values constrained by a previous spectral
analysis of these data  \citep{Keek2016}. For the spin, disk inner
radius, inclination angle, ionisation parameter, iron abundance, slope
of the incident power-law and high energy cut-off, we use $a=0.89$,
$r_{\rm in}=$ISCO, $i=30^\circ$, $\log_{10}\xi=2.68$, $A_{\rm
  Fe}=3.9$, $\Gamma=1.95$ and $(E_{\rm cut})_{\rm obs}=300$ keV (we
assume a constant ionisation parameter for simplicity). Following
\cite{Kalberla2005}, we fix $N_h=3.6 \times 10^{20} \, {\rm
  cm}^{-2}$. We use the model configuration that outputs time lags
accounting for line-of-sight absorption and the instrument response
(\texttt{ReIm}=6), and calculates $\phi_A$ self-consistently
(\texttt{PHI\_SET}=1). The remaining free parameters are black hole
mass $M$, source height $h$ and the boost parameter $1/\mathcal{B}$
(for which we set the hard ranges $>2~R_g$ and $0-3$ respectively).

The blue solid line represents our best-fitting model, which has
parameters $M=6.8^{+5.5}_{-5.9} \times 10^{6} M_{\odot}$, $h =
2.2^{+34}_{-p}$ and $1/ \mathcal{B} = 0.63^{+p}_{-0.48}$ ($\chi^2/$d.o.f. =
11.3/9; errors are $1\sigma$), where $p$ indicates that a
parameter is pegged at a hard limit. This value of mass is smaller
than the optical reverberation measurements in the literature, with
the two most recently published values being $M=[14.2 \pm 3.7] \times
10^6~M_\odot$ \citep{Peterson2004} and $M=[26 \pm 8] \times
10^6~M_\odot$ \citep{Grier2012}. \cite{Chainakun2016}  recently also
fit an X-ray reverberation model to the same lag spectrum and obtained
a best fitting mass of $13.5 \times 10^6~M_\odot$, albeit with poorly
constrained errors due to the computational expense of their model. We
investigate this apparent  discrepancy by re-fitting our model with
the mass fixed to the two optical reverberation values. The red dotted
and yellow dashed lines in Fig. \ref{fig:fit_mrk335} (left) show the
resulting best fits. The $M=14.2 \times 10^6~M_\odot$ fit has
parameters $h= 26.9$ and $1/\mathcal{B}=3$ ($\chi^2/$d.o.f. = 14.4/9)
and the $M=26 \times 10^6~M_\odot$ fit has parameters $h= 10.5$ and
$1/\mathcal{B}=3$ ($\chi^2/$d.o.f. = 13.7/9). We see that these two
high mass fits have very similar lag-energy spectra. We note that the
iron line feature in our model (and in the model of
\citealt{Chainakun2016}) is far less prominent than the Gaussian line
feature in the empirical model plotted in Fig 8 of
\cite{Kara2013}. However our best fitting model, which has one less
free parameter than their linear plus Gaussian model, provides a
better statistical description of the data (the empirical model has
$\chi^2$/d.o.f. = 13.22/8).

An F-test reveals that our best-fitting mass is preferred to the
\cite{Peterson2004}, \cite{Grier2012} and \cite{Chainakun2016} mass
values with only $\lesssim 1.5~\sigma$ confidence.
The reason why very different masses can give such simular $\chi^2$
values is particularly fascinating, and  serves to illustrate the
importance of fitting to multiple frequency ranges instead of just
one. Fig. \ref{fig:fit_mrk335} (right) shows the lag averaged over the
iron line energy range plotted against frequency for the three models (colours
and line styles have the same meaning as in the left hand plot). The
grey band denotes the frequency range used for our fits. We
see that, averaged over this frequency range, the three models 
give roughly the same time lag as each other ($\sim 10$ s). However,
the two high mass models diverge enormously from our best fitting
model at lower frequencies. We see that the high mass models are
actually in the phase-wrapping regime in the frequency range used for
the fit. This happens when the time lag
between the direct and reflected signals is greater than $\pi \nu$ or
less than $-\pi \nu$, similar to the effect that leads to car wheels
appearing to rotate backwards when viewed on film with a frame rate
lower than the rotation frequency of the wheels. Note that the time lag between the
two energy bands used for this figure is \textit{not} greater than
$\pi \nu$ when phase wrapping starts. This is because of
dilution: the time lag between \textit{direct} and \textit{reflected}
components is $>\pi\nu$, but both energy bands contain some direct and
some reflected X-rays (see the discussion in
\citealt{Uttley2014}). Fig \ref{fig:cont_mrk335} is a 2D $\chi^2$
  contour plot of black hole mass and source height that illustrates this
  point further. We see there are two dark stripes corresponding to
  regions of statistically acceptable mass (plus several lighter
  stripes in the top right hand corner). The two optical reverberation
  masses fall in the upper stripe (blue crosses), which therefore
  corresponds to our phase-wrapped regime. Our best fit falls in the
  lower stripe (green cross), and the error estimate quoted above of
  $\pm \sim 6\times 10^6~M_\odot$ only considers this lower stripe. We
  also see an anti-correlation between mass and source height. This
  occurs because the light-crossing time lag depends on $h \times
  R_g$, and therefore an increase in $h$ can be offset to some extent
  by a decrease in $M$.

At lower frequencies ($\lesssim 10^{-5}$ Hz), the $5.5-6.5$ keV reverberation lag for the high
mass models is \textit{far} larger than for the low mass model. This
is partly due to the larger mass itself (i.e. $1~R_g$ is a larger
distance), and partly because the source height and boost parameter
are both much larger in the high mass models (i.e. large $h$ means the
path-length difference is a greater number of gravitational radii, and
large $1/\mathcal{B}$ means that the reflection fraction is still high
even though $h$ is large, thus reducing dilution). This means that, in
the frequency range used for the fits, the phase-wrapped time lag in
the high mass models is roughly similar to the time lag in the low
mass model. The high mass models therefore predict that there should
be a negative iron K lag feature at $\nu\sim 10^{-4}$ Hz and a very
large positive lag at even lower frequencies, although
  these features will be heavily diluted by the intrinsic hard
  lags. Even though the lowest frequencies ($\nu \lesssim 10^{-5}$
Hz) cannot be probed with currently available data, it should be
possible to test these predictions in future by fitting a
  modified version of the model that additionally models hard 
  lags as fluctuations in the photon index for a number of frequency
ranges, yielding a robust mass measurement in the process. Of the
models we explore here, the low mass model is the more plausible, due
to the very high boost parameter (pegged to its hard upper limit)
required for the high mass models, although the parameters will
  likely change once hard lags are accounted for in this frequency
  range, which is necessary to also reproduce the observed
  variability amplitude. The results of \cite{Chainakun2016} instead
favour a higher mass, and their model included an ionization profile
and a simultaneous fit to the time-averaged spectrum. However, it is
not clear whether or not their best fit was in the phase wrapping
regime.

\section{Discussion}
\label{sec:discussion}

We have presented a public \textsc{xspec} reverberation mapping
model that can be fit to the energy dependent complex
cross-spectrum of black hole X-ray binaries and AGN for a range of
Fourier frequencies. It is now common to fit the time-averaged
spectrum with sophisticated relativistic reflection models. Our model
is designed to be comparably user-friendly to the spectral models, but
with the considerable extra functionality of also modelling the timing
properties. This provides the opportunity for better geometrical
constraints and entirely new black hole mass constraints.

\subsection{Comparison with previous work}

We have compared our model extensively to the existing spectral model
\textsc{relxilllp}, and find good agreement with the most recent
version of that model. We did however find a very minor error in the
\cite{Bardeen1972} expression for the Lorentz factor of a rotating
disk element, which has propagated into the \textsc{relxilllp} model
and likely somewhat further into the literature (see Appendix
\ref{sec:area}). However, we find the discrepancy is small enough to
be inconsequential. Further bench marking against
other spectral models (e.g. \citealt{Dovciak2004,Wilkins2012}) will be
very useful.

Previous reverberation mapping modelling studies have mainly focused
on AGN time lags
(\citealt{Cackett2014,Emmanoulopoulos2014,Chainakun2015,Caballero-Garcia2018}). Ours
is the first public model to also consider variability amplitudes. Our
model, similar to most previous studies, uses the lamppost
geometry. There has been work to model more sophisticated geometries
that self-consistently produce hard intrinsic lags through propagating mass
accretion rate fluctuations (\citealt{Wilkins2013,Wilkins2016}), but
these models are too computationally expensive for fitting to
data. The two blobs model of \cite{Chainakun2017}, consisting of two
lamppost sources, allows a slightly more realistic geometry that also
can produce hard intrinsic lags but without prohibitive computational
expense. Such a geometry can be used in our model, as long as the user
fits for real and imaginary parts of the cross-spectrum rather than
amplitude and time lags. In this case, two \textsc{reltrans} model
components with different source heights can simply be added
together. Intrinsic lags are then produced if the amplitude and phase
normalisations of the two components - $\alpha_1(\nu)$,
$\alpha_2(\nu)$, $\phi_{A1}(\nu)$ and  $\phi_{A1}(\nu)$ - are left as
free parameters. This essentially models a lag between incident
emission from the two lamppost sources, as in \cite{Chainakun2017}.

\subsection{Ionization profile}

We include a self-consistently calculated radial disk ionization
profile in our model and find that this has a significant effect on
the model outputs.
There is some uncertainty over the radial disk density profile that
should be used to calculate the ionization profile. Our model
considers both a Shakura-Sunyaev zone A (radiation pressure dominated)
density profile, and a constant density. Even assuming the
\cite{Shakura1973} model to be exact, we still expect the density
profile to depend on mass accretion rate, black hole mass and the
viscosity parameter. In particular, the disks of X-ray binaries in
faint hard states likely do not have a radiation pressure dominated 
zone, especially given the weight of evidence for disk truncation in
this state (e.g. \citealt{Tomsick2009,Ingram2017}). Interestingly
though, this implies that there will be a point in 
the outburst at which the mass accretion rate has risen sufficiently
for the inner disk to become radiation pressure dominated, leading to
a change in ionization profile. Perhaps with careful modelling, this
may be detectable with high quality data from current observatories
such as \nicer~and \nustar, or future observatories such as \athena,
\strobex~or \textit{Colibr\`i}
\citep{Ray2019,Caiazzo2019}. Constraining this transition would
provide useful insights  into disk physics, such as estimating the
viscosity parameter in the inner disk. Fitting the cross-spectrum for
a wide range of frequencies in addition to the time-averaged spectrum
will be far more constraining in this respect than only considering
the spectrum.

Although we account for the ionization profile, we do not account for
the dependence of the rest frame reflection spectrum on the density
itself (\citealt{Garcia2016}). We use the public \textsc{xillver} and
\textsc{xillverCp} grids, that are hardwired for $n_e=10^{15}$
cm$^{-3}$. This value is more appropriate for the most massive AGN
than for X-ray binaries, whose disks are expected to be much denser
($n_e \propto 1/M$, assuming the disk to be radiation pressure
dominated and in vertical hydrostatic equilibrium). The main
difference is a much higher disk temperature and therefore much more
thermal radiation in soft X-rays. The effect of radially stratified
density has not yet been explored.

When we generate fake data from a model with a self-consistent
ionization profile and fit with a constant ionization model, we find
that a narrow (non-relativistically smeared) reflection component is
required in the fit with high statistical significance (although we
note that a more systematic parameter exploration would be required to
make strong conclusions). This is interesting because fits to real
data commonly require such a narrow reflection component, which can be
attributed to a distant reflector
(e.g. \citealt{Garcia2015,Ingram2017}). This can either take the form
of a flared outer disk, the companion star for X-ray binaries, or the
torus for AGN. However, the flux of the best-fitting narrow reflection
component is often uncomfortably high. Our result suggests that these
high fluxes could actually be down to a modelling systematic that
could be at least in part alleviated by using an ionization
profile. The timing-properties are also very sensitive to the
ionization profile. Indeed \cite{Chainakun2016} found that the $\sim
3$ keV dip present in the lag spectrum of a number of AGN could only
be explained by stratification of the disk ionization
parameter. We also note that, whereas a parameter combination can be
found to allow a single ionization model to mimick the
time-averaged spectrum produced by a model with self-consistent
ionization, additionally considering the time lags and variability
amplitude should break the degeneracy.

We also found from our fits to fake data that a constant ionization
model under-predicts the source height (consistent with the results of
\citealt{Svoboda2012} and \citealt{Kammoun2019}). \cite{Niedzwiecki2016}
noted two problems associated with the very low source heights
measured by many spectral  fitting studies of AGN and X-ray binaries
(e.g. \citealt{Parker2014,Kara2015,Parker2015,Degenaar2015,Beuchert2017,Wang2017}).
First, the resulting surpression of the directly observed flux through
gravitational redshift and lensing means that fits to bright sources
such as Cygnus X-1 (e.g. \citealt{Parker2015,Beuchert2017}) require an
intrinsic source flux as high as $\sim 50$ times the Eddington limit
for a hard spectral state. Second, the intrinsic high energy cut-off
implied by such a large source redshift is so high that runaway
cooling should have long since been triggered by pair production
(e.g. \citealt{Poutanen1996,Fabian2012}). The higher source heights
yielded by accounting for a realistic ionization profile alleviate
both of these problems.

\subsection{Time lags and instrument response}

In Section \ref{sec:lagbias}, we show that failing to account for
line-of-sight absorption and the instrument response matrix can
significantly bias the predicted time lags. This bias is particularly
prominent in the $\lesssim 1$ keV energy range of \xmm~data, but has
little to no effect in the $\sim 2-10$ keV range. This may at least
partly explain why studies of AGN that model the $\sim 0.3-1$ vs $\sim
1-10$ keV lags (e.g. \citealt{Emmanoulopoulos2014}) have returned
lower source heights than those modeling the $\sim 5-7$ vs
$\sim 2-4$ keV lags (\citealt{Epitropakis2016}). This is because
ignoring the telescope response over-predicts the soft band lag for a
given source height (see Fig \ref{fig:abs}), and so a very small
source height is needed to still produce the fairly small lags present
in the data. For frequency ranges in which the intrinsic hard lags are
prominent, the response matrix bias can give rise to spurious features
in the observed $E\lesssim 1$ keV lag spectrum that look worryingly
like the features in X-ray binary data previously attributed to thermal
reverberation (e.g. \citealt{Uttley2011,DeMarco2017}). We conclude
that the observation of GX 339-4 from \cite{DeMarco2017} and
\cite{Mahmoud2019} that we investigate does indeed contain a signature
of thermal reverberation, but that the measured value of the lag may
have been heavily biased by failure to account for the instrument
response. \cite{Mahmoud2019} fit a transfer function model to the GX
339-4 data and measure a disk inner radius of $\sim 20~R_g$. However,
their model only accounts for the effective area curve of \xmm~and not
the redistribution matrix. Their inner radius value may therefore be
an over-estimate, since the intrinsic lags are likely shorter than
what is inferred from their analysis.

\subsection{Black hole mass}

Our proof-of-principle fit to the lag-energy spectrum of Mrk 335 in a
single frequency range (Section \ref{sec:Mrk335}) favours a black hole
mass of $\sim 7$ million $M_\odot$ to the optical reverberation values
in the literature of $\sim 14$ million $M_\odot$ \citep{Peterson2004}
and $\sim 26$ million $M_\odot$ \citep{Grier2012}, and the previous
X-ray reverberation value of $\sim 13.5$ million $M_\odot$
\citep{Chainakun2016}, although the higher masses are only disfavoured
with $\lesssim 1.5~\sigma$ confidence. The confidence range on the
mass is very large for such a fit to a single frequency range, partly
because the size of the reverberation lag is degenerate with the
reflection fraction. However, our findings here demonstrate very
effectively that this degeneracy will be eliminated by a simultaneous
fit to multiple frequency ranges, since our best-fitting $M =
6.8\times 10^6~M_\odot$ model predicts a wildly different time lag
signature at lower frequencies to the two higher mass models that we
also explore. For this it will be vital to additionally model
  the intrinsic hard lags. In fact, we find that intrinsic hard lags
  are required in order to explain the variability amplitude in addition
  to the lags even in the frequency range explored here. This may
  therefore bias the black hole mass yielded by our current analysis. We will
conduct a full multi-frequency analysis on the Mrk 335 data in
a future paper, also simultaneously considering the variability
amplitude and time-averaged spectrum (Mastroserio et al in prep). The
resulting constraints on  the black hole mass may enable some of the
uncertainties associated with optical reverberation mapping to be
addressed, most notably the uncertain geometry of the broad line 
region (typically parameterised by the constant $f$). We note that the
X-ray reverberation analysis of \cite{Emmanoulopoulos2014} returned a
black hole mass estimate of $M=19.8^{+11.8}_{-10.5} \times 
10^{6}~M_\odot$ for Mrk 335, which is again larger than our
value. Their fit procedure was very different to ours and that of
\cite{Chainakun2016}. They fit for time lags between two broad energy
bands as a function of frequency, and employed simplified assumptions
regarding the energy dependence of reflected X-rays.

X-ray reverberation mapping can also be used to measure the mass of
stellar-mass black holes in X-ray binary systems. In a companion
paper, we constrain the mass of the black hole in Cygnus X-1
(Mastroserio et al submitted). We do however note that care must be taken
to avoid frequency ranges dominated by quasi-periodic oscillations,
given strong evidence that these are driven by geometric oscillations
that are not modelled here (\citealt{Ingram2015,Ingram2016}). 

\subsection{Future modelling improvements}

Our model is still very idealised, and there is much room for future
improvement. We will in future extend our model to account for 
fluctuations in the power-law index of the illuminating spectrum
(Mastroserio et al submitted). It will also be useful in future to
include a non-zero disk scale height. \cite{Taylor2018} and
\cite{Taylor2018b} show that using the scale height expected for a
radiation pressure dominated disk leads to steeper emissivity
profiles, with spectral and timing properties consequently adjusted
due to the signal being more dominated by the broader line and shorter
time lags associated with the inner regions of the disk. Our
calculation also assumes that the source is stationary, whereas
emission would actually be boosted
somewhat away from the disk if the source is actually a standing shock
at the base of the outflowing jet, as is often suggested
(\citealt{Markoff2005,Dauser2013}). We do however include a `boosting
parameter' that approximates this affect by reducing the reflection
fraction. If the user takes the lamppost geometry seriously and finds
that the boosting parameter is statistically required in the fit to be
less than unity, they can conclude that the source is moving away from
the disk. \cite{Niedzwiecki2018} recently pointed out that in a
lamppost geometry, we should also sometimes see the other lamppost
source on the underside of the disk and we should also see photons
from the top lamppost who's trajectories have bent around the black
hole and into our line of sight - particularly if the disk is
truncated. We do not include these effects, which will
presumably be blocked or significantly altered by material inside of
the disk.

We use the models \textsc{xillver} and \textsc{xillverCp} to compute
the rest frame reflection spectrum (\citealt{Garcia2010,Garcia2013}),
which are state-of-the-art, but still include approximations that can
be addressed in future. A constant vertical density profile is
assumed, which returns a very different $E\lesssim 1$ keV reflection
spectrum from a calculation assuming vertical hydrostatic equilibrium
(\citealt{Nayakshin2000,Done2007a,Rozanska2011,Vincent2016}). We note,
however, that recent numerical simulations indicate that the vertical
density profile of a magnetic pressure dominated disk is roughly
constant near the surface \citep{Jiang2019}. The largest approximation
of all is likely the lamppost model itself, and so it will be important to
explore more realistic geometries in future \citep{Zhang2019}.

\section{Conclusions}
\label{sec:conclusions}

The X-ray reverberation models \textsc{reltrans} and
\textsc{reltransCp} are now publicly available for use in
\textsc{xspec}. The source code and usage instructions can be
downloaded from \url{https://adingram.bitbucket.io/}. The models can
be used to simultaneously fit the real and imaginary parts of the
energy-dependent cross-spectrum for a wide range of Fourier
frequencies, plus the time-averaged spectrum. Intrinsic hard lags can
be accounted for by using two model components added together. The
model is designed to be user friendly for the beginner but flexible
for the advanced user, with environment variables specifying model
properties and advanced options. We find that modelling systematics
have likely led to artificially low source heights being measured in
the literature. We also find that bright distant reflection component
often statistically required in spectral fits can at least partially
be explained by the radial profile of the disk ionization
parameter. Our proof-of-principle fits to the lag-energy spectrum of
the Seyfert galaxy Mrk 335 return a smaller mass for the central black
hole than previous optical reverberation mapping analyses ($\sim 7$
milion compared with $\sim 14-26$ million $M_\odot$), which we will
investigate in more detail in future.

\section*{Acknowledgements}

A. I. acknowledges support from the Royal Society and valuable
discussions with Chris Reynolds and George Ellis. G.M. acknowledges
support from NWO.



\appendix

\begin{onecolumn}

\section{Area of a disk ring}
\label{sec:area}

The proper area of a disk annulus of width $dr$ as measured by a
stationary observer is given by $d^2x = 2\pi \sqrt{ g_{rr}
  g_{\phi\phi} } dr$ (see e.g. \citealt{Wilkins2012}). The disk area
element we are after for our calculation is measured in the disk
frame, bringing in a factor of $\gamma^\phi$, which is the Lorentz
factor of the orbiting disk element. Substituting in the components of
the Kerr metric gives
\begin{equation}
\frac{dA_{\rm ring}}{dr} = 2\pi \gamma^\phi \sqrt{ \frac{r^4+a^2 r^2 + 2 a^2 r}{
  r^2 - 2r + a^2 }   }.
\end{equation}
This equation agrees with the formula derived by \cite{Wilkins2012}
and \cite{Dauser2013}, except for a small typographical error in
\cite{Dauser2013}. We see that, in the limit $r>>2$, this reduces to
$2\pi r$, as we would expect.

Formulae for the Lorentz factor are presented in \cite{Bardeen1972}
(hereafter BPT72) and \cite{Dauser2013}. However, a very small error in
BPT72 has propagated into the later literature. We therefore present a
derivation here. In order to do this, we must first define a local
non-rotating frame (LNRF) in which $r=$constant, $\theta=$constant and
$\phi=\omega t +$constant. Here, $\omega=-g_{t\phi}/g_{\phi\phi}$ is
the term that allows the reference frame to rotate with inertial
frames (i.e. the frame dragging effect). For any stationary,
axisymmetric, asymptotically flat spacetime, we can write the line
element $ds^2 = g_{\mu\nu}dx^\mu dx^\nu$ as
\begin{equation}
ds^2 = -{\rm e}^{2\nu} dt^2 + {\rm e}^{2\psi} (d\phi - \omega dt)^2 +
{\rm e}^{2\mu_1} dr^2 + {\rm e}^{2\mu_2} d\theta^2,
\end{equation}
where the exponentials are defined in equations 2.3 and 2.5 of BPT72 for the
case of the Kerr metric. Setting $M=1$ in the BPT72 equations gives
the dimensionless units we employ here.

We can represent the 4-velocity in the LNRF as $u^{(a)} = u^\mu
e_\mu^{(a)}$, where the components of the tetrad of basis vectors are
given by $e^{(i)} = e_\mu^{(i)} dx^\mu$. The (covariant) tetrad of
basis vectors for the LNRF is (BPT72 equation 3.2)
\begin{equation}
e^{(t)} = {\rm e}^\nu dt,~~~~ {\rm e}^{(r)} = {\rm e}^{\mu_1} dr,~~~~
{\rm e}^{(\theta)} =
{\rm e}^{\mu_2} d\theta,~~~~ {\rm e}^{(\phi)} = -\omega {\rm e}^\psi
dt + {\rm e}^\psi d\phi,
\end{equation}
where it is a long-standing travesty that the letter e is used both
for the basis vectors and as the exponential number (we try to clear
this up by using an italicised font for the basis vectors). This gives
\begin{equation}
\mathbf{e^{(t)}}_\mu = ({\rm e}^\nu,0,0,0) ,~~~~ \mathbf{e^{(r)}}_\mu = (0,{\rm e}^{\mu_1},0,0)
,~~~~ \mathbf{e^{(\theta)}}_\mu = (0,0,{\rm e}^{\mu_2},0)
,~~~~ \mathbf{e^{(\phi)}}_\mu = (-\omega {\rm}^\psi,0,0,{\rm e}^{\psi}).
\label{eqn:tetrad}
\end{equation}
The covariant tetrad can be derived from the definition $e_\mu^{(a)}
e_\nu^{(b)} g^{\mu\nu} = \eta^{(a)(b)}$, and the contravariant tetrad
from $e_{(a)}^\mu e_{(b)}^\nu g_{\mu\nu} = \eta_{(a)(b)}$, where
$\eta$ is the Minkowski metric.

The 3-velocity is $v^{(i)} = u^{(i)} / u^t$. For circular, equatorial
orbits, the only non-zero component of the 3-velocity is the $\phi$
component, which is given by
\begin{equation}
v^{(\phi)} = (\Omega_\phi-\omega) {\rm e}^{\psi-\nu}
\label{eqn:vphi}
\end{equation}
where $\Omega_\phi=u^\phi/u^t$ is the angular velocity of the disk
element. In the Kerr metric, this is $\Omega_\phi = \pm 1/(r^{3/2} \pm a)$,
where the top and bottom signs are respectively for prograde and
retrograde spin. Equation (\ref{eqn:vphi}) is the same as equation
3.10 in BPT72 except for a small typographical error in the index of
the exponential in the BPT72 version. Subbing in the Kerr metric gives
\begin{equation}
v^{(\phi)} = \frac{ r^2 + a^2 - 2 a r^{1/2} + 2 r^{-1}(a^2\pm a^2) }{
  \Delta^{1/2} (r^{3/2} \pm a) },
\label{eqn:vphi2}
\end{equation}
where $\Delta = r^2 - 2 r + a^2$. The Lorentz factor is then simply
given by $\gamma^\phi = [ 1 - (v^{(\phi)})^2 ]^{-1/2}$. Equation
(\ref{eqn:vphi2}) agrees with equation 3.11a in BPT72 for prograde
spin but not for retrograde. Equation 10 in \cite{Dauser2013} can be
reproduced by taking equation 3.11a from BPT72 and dropping the $\pm$
and $\mp$ signs. Therefore, our equation (\ref{eqn:vphi2}) is valid
for prograde and retrograde spins, whereas the equivalent equations
from BPT72 and \cite{Dauser2013} (and potentially many other
references) are only strictly accurate for prograde spin. In
practice, the inaccuracy introduced by these mistakes is very small
and need not be worried about.

\section{Blueshift factors and angles}
\label{sec:gandmu}

Here we present the formulae used to calculate the various blueshift
factors and angles. The covariant form of the tangent 4-vector of
photons following geodesics in the Kerr metric is
(see e.g. \citealt{Bardeen1972,Dauser2013})
\begin{equation}
(\mathbf{k})_\mu = (-1,\pm\sqrt{V_r}/\Delta,\pm|q|,0),
\end{equation}
where $q$ is Carter's constant \citep{Carter1968} and
\begin{equation}
\Delta = r^2-2r+a^2;~~~~~~~V_r=(r^2+a^2)^2 - \Delta (q^2+a^2).
\end{equation}
Carter's constant for `incident' photons propagating from source to
disk is (e.g. \citealt{Dovciak2004})
\begin{equation}
q_i^2 = \frac{\sin\delta (h^2+a^2)^2}{h^2-2h+a^2}-a^2,
\end{equation}
and Carter's constant for `emergent' photons propagating from a disk
element to the observer is (e.g. \citealt{Dovciak2004,Ingram2015a})
\begin{equation}
q_e^2 =\beta^2 + \cos^2i~(\alpha^2-a^2).
\end{equation}
4-velocity can be expressed as
\begin{equation}
\mathbf{u}^\mu = \frac{ \mathbf{\Omega}^\mu }{ \sqrt{-g_{\alpha\beta} \Omega^\alpha
    \Omega^\beta } },~~~{\rm where}~~~\mathbf{\Omega}^\mu \equiv \frac{d\mathbf{x}^\mu}{dt}.
\end{equation}
The 4-velocity of the disk element is therefore
\begin{equation}
(\mathbf{u_d})^\mu = u_d^t ( 1 , 0 , 0, \Omega_\phi
);~~~~~~~\Omega_\phi=\frac{\pm 1}{r^{3/2} \pm a};~~~~~~~u_d^t = \left\{ 1 - \frac{2}{r}
  + \frac{4 a \Omega_\phi}{r} - \left[ r^2 +a^2 \left( 1+\frac{2}{r} \right) \right] \Omega_\phi^2 \right\}^{-1/2},
\end{equation}
and the 4-velocity of the stationary source is
\begin{equation}
(\mathbf{u_s})^\mu =\frac{1}{ \sqrt{ -g_{tt} } } ( 1 , 0 , 0, 0 ) =
\sqrt{  \frac{ h^2+a^2 } { h^2 - 2h + a^2 } } ( 1 , 0 , 0, 0 ).
\end{equation}
The blueshift seen by an observer on the disk patch is therefore
\begin{equation}
g_{sd}(r) = \frac{ (k_i)_\mu (u_d)^\mu }{
  (k_i)_\nu (u_s)^\nu } = \frac{ (u_d)^t }{ (u_s)^t
} = \sqrt{\frac{h^2-2h+a^2}{h^2+a^2}} \left\{ 1 - \frac{2}{r}
  + \frac{4 a \Omega_\phi}{r} - \left[ r^2 +a^2 \left( 1+\frac{2}{r} \right) \right] \Omega_\phi^2 \right\}^{-1/2}.
\end{equation}
It follows that the blueshift experienced by photons propagating from
the stationary lamppost source to a distant stationary observer is
\begin{equation}
g_{so} = \frac{1}{u_s^t} = \sqrt{\frac{h^2-2h+a^2}{h^2+a^2}}.
\end{equation}
The cosine of the incidence angle is given by
\begin{equation}
\mu_i = - \frac{ (k_i)_\mu e^\mu_{(\theta)} } { (k_i)_\nu (u_d)^\nu }
= \frac{ (k_i)_\theta e^\theta_{(\theta)} } { (u_d)^t } = \frac{ (q_i/r)
} { (u_d)^t },
\end{equation}
because the $\theta$ component of the contravariant tetrad of basis
vectors is $e^\theta_{(\theta)}=1/r$. For the blueshift experienced by
emergent photons propagating from disk element to observer,
$g_{do}(r,\phi)$, we use equation (4) from \cite{Ingram2017}. This is
accurate for a razor thin disk in the black hole equatorial plane. The
cosine of the emission angle is 
\begin{equation}
\mu_e = \frac{(k_e)_\mu e_{(\theta)}^\mu }{ (k_e)_\nu (u_d)^\nu  } =
g_{do}(r,\phi) (k_e)_\theta e_{(\theta)}^\theta = g_{do}(r,\phi) \frac{q_e}{r}.
\end{equation}

\section{Environment variables}
\label{sec:env}

The environment variables used by the model as listed in Table
\ref{tab:env}. All have sensible default values that the user can
override, for example to change the model resolution or explore
different radial ionization profiles.

\begin{table*}
\begin{center}
\begin{tabular}{|p{3cm}|p{9.5cm}|p{2cm}|p{1.5cm}|} 
\hline
Name  & Function & Possible values & Default value \\
\hline
\hline
\texttt{REV\_VERB} & Sets the verbose level. & 0-1 & 0 \\
\hline
\texttt{PHI\_SET}   & Sets whether $\phi_A$ is calculated self-consistently (1)
             using equation \ref{eqn:phiA} or set by the model
             parameter \texttt{phiA} (0). & 0-1 & 0 \\
\hline
\texttt{RMF\_SET}   & Name (including path) of the instrument response file. If
             the code is in a mode requiring an instrument response,
             the user is prompted for this. & string & none \\
\hline
\texttt{ARF\_SET}   & Name (including path) of the instrument ancillary
             response file. This is not required if a .rsp file is
             entered. & string & none \\
\hline
\texttt{MU\_ZONES}   & Sets how many bins of $\mu_e(r,\phi)$ are used for the
              calculation. If this is set to 1, $\mu_e=\cos i$ is
              assumed. & 1-10 & 5 \\
\hline
\texttt{ECUT\_ZONES}   & Sets how many bins of $g_{sd}(r)$ are used for the
              calculation. If this is set to 1, it is assumed that the
                high energy cut-off seen by each disk ring is equal to
                $E_{cut}$ rather than $g_{sd}(r) E_{cut}$. & 1-10 & 5 \\
\hline
\texttt{ION\_ZONES}   & Sets how many bins of $\log_{10}\xi(r)$ are used for the
              calculation. If this is set to 1, the ionization is
               assumed to be constant and determined by the input
               model parameter \texttt{logxi}. Otherwise, a radial
               ionization profile is used, with \texttt{logxi} setting
               the normalisation of the profile. & 1-100 & 10 \\
\hline
\texttt{A\_DENSITY}   & Sets whether the radial ionization profile is
               calculated assuming a constant disk density (0), or
               assuming a zone A Shakura-Sunyaev disk density profile
               (1). The parameter \texttt{logxi} respectively sets
               exactly or roughly the peak value of $\log_{10}\xi(r)$
               for the former and latter cases. & 0-1 & 1 \\
\hline
\texttt{GRID}   & Name of geodesics grid (with full path). If a value
                  is given, the grid is used to calculate an
                  interpolated transfer function. Otherwise, all geodesics
                 are calculated on the fly. Use of the grid is faster
                  if $a$ and $i$ are free parameters, and the on the
                  fly calculation is faster if they are fixed.  & string  & null \\
\hline
\hline
\texttt{RELXILL\_TABLE\_PATH}   & \textsc{relxill} variable: points to the
                         directory containing the \textsc{xillver}
                         tables. & string & null \\
\hline
\end{tabular}
\end{center}
\caption{Environment variables. The default variables are used if the
  user does not set any of the above variables.}
\label{tab:env}
\end{table*}


\end{onecolumn}

\bsp	
\label{lastpage}
\end{document}